\documentclass{aa}  

\usepackage{graphicx}
\usepackage{txfonts}

\usepackage[breaklinks,colorlinks,citecolor=blue]{hyperref}
\usepackage{ulem}

\usepackage{lmodern}

\bibpunct{(}{)}{;}{a}{}{,} 

\newif\iffigs
\newif\iffigscl
\newif\iffigstest
\newif\iflabs

\figstrue

\figscltrue

\figstestfalse

\labsfalse

\begin{document} 


\newif\iffigs

\figstrue

\newcommand{\itii}[1]{{#1}}
\newcommand{\franta}[1]{\textbf{\color{green} #1}}
\newcommand{\frantaii}[1]{\textbf{\color{magenta} #1}}
\newcommand{\itiitext}[1]{{#1}}

\newcommand{\eq}[1]{eq. (\ref{#1})}
\newcommand{\eqp}[1]{(eq. \ref{#1})}
\newcommand{\eqq}[1]{eq. \ref{#1}}
\newcommand{\eqb}[2]{eq. (\ref{#1}) and eq. (\ref{#2})}
\newcommand{\eqc}[3]{eq. (\ref{#1}), eq. (\ref{#2}) and eq. (\ref{#3})}
\newcommand{\refs}[1]{Sect. \ref{#1}}
\newcommand{\reff}[1]{Fig. \ref{#1}}
\newcommand{\reft}[1]{Table \ref{#1}}

\newcommand{\datum} [1] { \noindent \\#1: \\}
\newcommand{\pol}[1]{\vspace{2mm} \noindent \\ \textbf{#1} \\}
\newcommand{\code}[1]{\texttt{#1}}
\newcommand{\figpan}[1]{{\sc {#1}}}

\newcommand{\nbdvi}{\textsc{nbody6} }
\newcommand{\nbdvid}{\textsc{nbody6}}
\newcommand{\flash}{\textsc{flash} }
\newcommand{\flashd}{\textsc{flash}}
\newcommand{\sfe}{\mathrm{SFE}}
\newcommand{\mum}{$\; \mu \mathrm{m} \;$}
\newcommand{\rop}{$\rho$ Oph }
\newcommand{\HT}{$\mathrm{H}_2$}
\newcommand{\Halpha}{$\mathrm{H}\alpha \;$}
\newcommand{\HI}{H {\sc i} }
\newcommand{\HII}{H {\sc ii} }
\renewcommand{\deg}{$^\circ$}

\newcommand{\dd}{\mathrm{d}}
\newcommand{\acosh}{\mathrm{acosh}}
\newcommand{\sign}{\mathrm{sign}}
\newcommand{\cex}{\mathbf{e}_{x}}
\newcommand{\cey}{\mathbf{e}_{y}}
\newcommand{\cez}{\mathbf{e}_{z}}
\newcommand{\cer}{\mathbf{e}_{r}}
\newcommand{\ceR}{\mathbf{e}_{R}}

\newcommand{\llg}[1]{\log_{10}#1}
\newcommand{\pder}[2]{\frac{\partial #1}{\partial #2}}
\newcommand{\pderrow}[2]{\partial #1/\partial #2}
\newcommand{\nder}[2]{\frac{\dd #1}{\dd #2}}
\newcommand{\nderrow}[2]{{\dd #1}/{\dd #2}}

\newcommand{\Cmiii}{\, \mathrm{cm}^{-3}}
\newcommand{\Gcmii}{\, \mathrm{g} \, \, \mathrm{cm}^{-2}}
\newcommand{\Gcmiii}{\, \mathrm{g} \, \, \mathrm{cm}^{-3}}
\newcommand{\Kms}{\, \mathrm{km} \, \, \mathrm{s}^{-1}}
\newcommand{\Si}{\, \mathrm{s}^{-1}}
\newcommand{\Esi}{\, \mathrm{erg} \, \, \mathrm{s}^{-1}}
\newcommand{\Ee}{\, \mathrm{erg}}
\newcommand{\Yr}{\, \mathrm{yr}}
\newcommand{\Kyr}{\, \mathrm{kyr}}
\newcommand{\Myr}{\, \mathrm{Myr}}
\newcommand{\Gyr}{\, \mathrm{Gyr}}
\newcommand{\Msun}{\, \mathrm{M}_{\odot}}
\newcommand{\Rsun}{\, \mathrm{R}_{\odot}}
\newcommand{\Pc}{\, \mathrm{pc}}
\newcommand{\Kpc}{\, \mathrm{kpc}}
\newcommand{\Mpc}{\, \mathrm{Mpc}}
\newcommand{\Sd}{\Msun \, \Pc^{-2}}
\newcommand{\Ev}{\, \mathrm{eV}}
\newcommand{\Kk}{\, \mathrm{K}}
\newcommand{\Au}{\, \mathrm{AU}}
\newcommand{\Mas}{\, \mu \mathrm{as}}

\newcommand{\logP}[1]{$\rm{log}_{10}(P[\rm{days}]) #1$}
\newcommand{\pcci}{p_{\rm CC} (M_{\rm ecl})}
\newcommand{\pccii}{p_{\rm CC} (M_{\rm ecl}, t)}
\newcommand{\pcciii}{p_{\rm CC} (m_{\rm Ceph})}

   \title{On the dynamical evolution of Cepheids in star clusters}

   \authorrunning{Dinnbier, Anderson, and Kroupa}
   \titlerunning{Dynamical Evolution of Cepheids in Star Clusters}

   \author{Franti\v{s}ek Dinnbier\inst{1},
          Richard I. Anderson\inst{2}
          \and
          Pavel Kroupa\inst{1,3}
          }

   \institute{
             Astronomical Institute, Faculty of Mathematics and Physics, Charles University in Prague, V Hole\v{s}ovi\v{c}k\'{a}ch 2, 180 00 Praha 8, Czech Republic \\
             \email{dinnbier@sirrah.troja.mff.cuni.cz} \\
             \email{pavel.kroupa@mff.cuni.cz}
         \and
              Institute of Physics, Laboratory of Astrophysics, \'Ecole Polytechnique F\'ed\'erale de Lausanne (EPFL), Observatoire de Sauverny, 1290 Versoix, Switzerland\\
              \email{richard.anderson@epfl.ch}
         \and
             Helmholtz-Institut f\"{u}r Strahlen- und Kernphysik, University of Bonn, Nussallee 14-16, 53115 Bonn, Germany
             }

   \date{Received \today; accepted ??}

\abstract{
We investigate the occurrence of classical (type-I) Cepheid variable stars (henceforth: Cepheids) in dynamically evolving star clusters 
from birth to an age of approximately 300\,Myr. 
The clusters are modelled by the Aarseth code \nbdvid, and they feature a realistic stellar initial mass function 
and initial binary star population, single star and binary star evolution, 
expulsion of the primordial gas, and the tidal field of the galaxy. 
Our simulations provide the first detailed dynamical picture of how frequently Cepheids remain gravitationally bound to their birth clusters versus 
how frequently they occur in the field. 
They allow us to quantify the relevance of various cluster escape mechanisms and how they depend on stellar mass. 
Overall, the simulations agree with the empirical picture that a small fraction ($\approx 10\%$) of Cepheids reside in clusters, that cluster 
halo membership is relatively common, and that the majority of Cepheid hosting clusters have only a single Cepheid member. 
Additionally, the simulations predict that
a) Cepheid progenitors are much more likely to escape from low-mass than higher-mass clusters;
b) higher-mass (long-period) Cepheids are $\approx 30\%$ more likely to be found in clusters than low-mass (short-period) Cepheids;
c) the clustered Cepheid fraction increases with galactocentric radius since cluster dispersal is less efficient at greater radii;
d) a lower metallicity reduces the overall clustered Cepheid fraction because the lower minimum mass 
of Cepheids leaves more time for cluster dispersal (this primarily affects short-period Cepheids);
e) high-mass clusters are much more likely to have more than one Cepheid member at any given time, in particular at a lower metallicity.
We interpret the results as outcomes of various aspects of star cluster dynamics. 
The comparison of predicted and observed clustered Cepheid fractions, $f_{\rm CC}$, highlights the need for additional cluster disruption mechanisms, 
most likely encounters with giant molecular clouds, to explain the observed fractions.
}
 


   \maketitle
%

\section{Introduction}

Classical Cepheids occurring in star clusters (henceforth: clustered Cepheids) are doubly important objects for stellar physics and cosmology (via the distance scale) because both their pulsations and their host cluster's member stars provide stringent tests of stellar models and allow to calibrate the Leavitt law \citep[LL, period-luminosity relation]{Leavitt1912}. The identification of Cepheids in clusters was initially a one-by-one procedure \citep{Irwin1955,Kholopov1956,Sandage1958} and required painstaking reddening corrections \citep[e.g.][and references therein]{Turner1976,Turner2002} before large amounts of high-quality astrometric data became available. 
\citet{Anderson2013} adopted a new strategy for  systematically  and quantitatively establishing the membership of Cepheids in Milky Way (MW) open star clusters using eight-dimensional phase-space information with an eye on the {\it Gaia} mission launched in December 2013 \citep{Gaiac2016b}. {\it Gaia}'s unprecedented astrometric census greatly facilitates the identification of MW star clusters based chiefly on proper motion clustering \citep[e.g.][]{CantatGaudin2018,Castro-Ginard2020}, while  ground- and space-based time-domain surveys \citep[e.g.][]{ASASSNvarstar2019,Soszynski2020galcep,GaiaDR2variability,Ripepi2019} considerably increased the number of MW Cepheid candidates. This combination has led to reassessments of cluster membership as well as several new interesting cluster Cepheid candidates \citep{Medina2021,Zhou2021}. However, a main limitation of such modern astro-statistical approaches is the formulation of prior probabilities based on (2D) on-sky location. Understanding how and how frequently Cepheids occur in star clusters or their halos would significantly help to improve the reliability of such priors. 

The overall empirical result of such studies has been that a) the vast majority of Cepheids do not reside in clusters and the clustered Cepheid fraction $f_{CC,\rm{true}} \lesssim 10\%$ \citep{Anderson2018}; b) MW clusters that do host Cepheids generally contain a single, occasionally two \citep[e.g.][]{Dekany2015}, and at the very most three Cepheids \linebreak \citep[NGC\,7790]{Majaess2013,Anderson2013,Medina2021}; c) a sizeable fraction of reported MW cluster Cepheids are located in the distant outskirts (halos) of the putative host clusters \citep{Anderson2013,Zhou2021,Medina2021}; d) $f_{CC,\rm{true}}$ is generally higher for the younger long-period Cepheids \citep{Anderson2018}. The most significant exceptions to this rule are very massive ($\approx 5 \times 10^4 \,\Msun$) clusters in the Large Magellanic Cloud, such as NGC\,1866 or NGC\,2031, that each contains up to 24 Cepheids, most of them with a fairly short-period \citep{Testa2007,Musella2016}. 


The identification of clustered Cepheids becomes increasingly difficult at distances beyond 100\,kpc both because the detection of the main sequence in the clusters becomes more difficult and because the angular resolution element of a detector corresponds to increasingly larger physical scales. Blending of Cepheids with cluster member stars depresses observed photometric amplitudes and alters light curve shapes, thus precluding the detection of Cepheids occurring in massive clusters \citep{Ferrarese2000,Riess2020}. This results in a lower observed \emph{effective} clustered extragalactic Cepheid fraction, $f_{CC,\rm{eff}}$, compared to the true number of clustered extragalactic Cepheids, and any uncertainties in $f_{CC,\rm{eff}}$ affect stellar association bias corrections for the distance ladder linearly \citep{Anderson2018}. Understanding the difference between $f_{CC,\rm{eff}}$ and $f_{CC,\rm{true}}$ and possible variations in $f_{\rm{CC}}$ among different galaxies will be very valuable for improving distance measurements based on Cepheids \citep{Anderson2018}. Providing a better astrophysical understanding of how Cepheids relate to their (former) host clusters further aids the calibration of the LL, which can be significantly improved by adopting precise average cluster parallaxes for LL calibration \citep{Breuval2020}.

The occurrence of clustered Cepheids is intimately related to star cluster dynamics because most stars are thought to be born in young embedded clusters \citep{Lada1991,Kroupa1995b,Lada2003,Porras2003,Megeath2016} that evolve dynamically over time, maintaining some imprints of the initial dynamical interactions within their dense birth environments. 
Following star cluster formation, these objects relax and mass segregate \citep[e.g.][]{Spitzer1969,Bonnell1998,Mouri2002,McMillan2007,Spera2016}, 
eject stars either in close interactions of three and more stars \citep[e.g.][]{Aarseth1971,Heggie1975,Perets2012,Tanikawa2012,Oh2015}
or as a result of a decay of primordial triples and quadruples \citep[e.g.][]{Sterzik1998,Goodwin2005,PflammAltenburg2006,Wang2019}, 
and evaporate stars as the result of weak encounters between stars \citep[e.g.][]{Kupper2008,Kupper2010}. 
The loss of stars from clusters is enhanced by the tidal field of the cluster hosting galaxy \citep{Baumgardt2002,Baumgardt2003,Sollima2020} and 
by removal of the residual gas which was not consumed in the star forming process \citep[e.g.][]{Lada1984,Kroupa2001b,Baumgardt2007}. 
The remaining fraction of stars that resides within their birth clusters then decreases as a function of age (assuming a constant star formation rate), as is observed for clustered Cepheids in the Milky Way, LMC, SMC, and M31 \citep[their Fig. 13]{Anderson2018}. 
Such a dynamical picture also explains the occurrence and aids the discovery of extended tidal tails of clusters dissolving into the MW field population \citep[e.g.][]{Roser2019a,Meingast2019,Jerabkova2021}. It may also help to confront predicted period-age relations \citep{Bono2005,Anderson2016} to observations \citep{Senchyna2015,Medina2021} because the dynamical timescale is independent of the assumptions or input physics of stellar models. This is particularly interesting, since Cepheid period-age relations are important to trace Galactic evolution even though the ages of Cepheids depend by nearly a factor of two on the effects of main sequence rotation \citep{Anderson2016}.

To address the above issues, we have computed an extensive library of $N$-body cluster simulations to investigate cluster dispersal over a timescale of up to $300$ Myr. In this first paper of a series exploiting these simulations, we present the computed library and its predictions concerning the occurrence of Cepheids in star clusters for different cluster masses, chemical composition, and galactocentric radius. 
Of course, cluster dynamics also impacts the orbital parameters of primordial binary stars \citep{Kroupa1995a,Oh2016}, and Cepheids are particularly likely to have experienced binary interactions, since the close orbital configurations  observed among their progenitors (B-type main sequence stars) must adapt to the significant evolutionary increase in radius (typical Cepheid radii $20-200$ R$_\odot$). 
Elucidating the nature and frequency of these binary interactions is crucial for forward modeling populations of Cepheids and for understanding the intriguing hierarchy of multiple star systems hosting Cepheids \citep{Evans2020}. Paper\,II of this series will focus on such effects (Dinnbier, Anderson, and Kroupa, in prep.).

This \emph{article} is structured as follows.
\refs{sec:simulations} describes the set up of the simulations, initial conditions, and other choices made to create the cluster library.
\refs{sIdentification} presents how Cepheids are identified and traced in the NBODY6 simulations.
\refs{sDynamical} investigates the evolution and occurrence of Cepheids in dynamically evolving star clusters as 
a function of the cluster mass, metallicity and galactocentric radius.
\refs{sFracCeph} seeks to estimate the number of Cepheids occurring in clusters, $f_{CC,\mathrm{true}}$, 
and establishes upper and lower limits based on the N-body simulations. 
The discussion in \refs{sDiscussion} focuses on the impact of gas expulsion, initial conditions, and the role of molecular clouds. 
The final \refs{sSummary}  summarizes our results and conclusions.

\section{Numerical methods and initial conditions}\label{sec:simulations}

\label{sInitCond}

\subsection{Numerics}

\label{ssNumerics}

The cluster simulations presented in this work are calculated by use of the Aarseth code \nbdvi \citep{Aarseth1999,Aarseth2003}. 
The code integrates stellar trajectories with a 4th order Hermite predictor-corrector scheme \citep{Makino1991}
with individual block time-steps coupled to the Ahmad-Cohen method \citep{Ahmad1973,Makino1992}. 
Stellar evolution is treated by stellar evolutionary tracks due to \citet{Tout1996} and \citet{Hurley2000}. 
To tackle the enormous range of relevant time-scales, \nbdvi employs various sophisticated numerical tools such as 
two-body \citep{Kustaanheimo1965}, three-body \citep{Aarseth1974a} and multiple-body \citep{Mikkola1990} regularisation, 
which include the dynamical formation of these systems, exchanges, and ionisations. 
The regularisation techniques typically transform the coordinates to a four-dimensional space, and they utilise the properties of the Levi-Civita matrix 
\citep{Kustaanheimo1965,Bettis1971,Stiefel1975}. 
The code initiates or terminates a particular regularisation technique according to given criteria. 
Not all compact systems are necessarily regularised. 
For example, unperturbed hierarchical systems are evolved as single bodies until an external perturbation or stellar evolution within the 
hierarchy necessitates an update of the system internal conditions \citep{Mardling1999}. 
During regularisation, the code approximates binary star evolution with Roche-lobe mass transfer \citep{Eggleton1983,Hurley2002}.
Stellar evolution produces compact objects, and it also handles the products of stellar mergers \citep{Kochanek1992}. 
We refer to \citet{Aarseth2003} for an in-depth description of relevant algorithms. 

We assume that all stars are formed in embedded star clusters containing a 
substantial reservoir of unprocessed gas at the time of massive star formation. 
The gas gets expelled due to feedback from massive stars, and the sudden change of the gravitational potential 
of the expelled gas affects the dynamics of the star cluster, which expands and loses a 
substantial fraction of its stars \citep{Lada1984,Goodwin1997,Geyer2001,Kroupa2001b}. 
To limit the number of free parameters, we adopt the gas expulsion model 
by \citet{Kroupa2001b}, where the gravitational
potential $\phi_{\rm gas}(r,t)$ of the gas is represented by a Plummer sphere that starts to exponentially decrease after a delay time $t_{\rm d}$, i.e. 
\begin{equation}
\phi_{\rm gas}(r,t) = 
\begin{cases}
-\frac{G M_{\rm gas}}{\sqrt{a_{\rm gas}^2 + r^2}} \;\; t \leq t_{\rm d}, \\
-\frac{G M_{\rm gas}}{\sqrt{a_{\rm gas}^2 + r^2}} \exp{\{-(t - t_{\rm d})/\tau_{\rm M}\}} \;\; t > t_{\rm d}.
\end{cases}
\label{ePotGas}
\end{equation}
In the gas model, $G$, $M_{\rm gas}$, $a_{\rm gas}$, and $\tau_{\rm M}$ denote the gravitational constant, 
initial mass of the gas component of the embedded star cluster, the Plummer length parameter of the gas component, and 
the gas expulsion time-scale, respectively. 
For the parameters of the gas expulsion model, we adopt the values which seem to be the most promising 
according to current understanding of embedded star clusters \citep{Kroupa2001b,Lada2003,Kuhn2014,Banerjee2017}. 
We adopt $a_{\rm gas} = a_{\rm ecl}$, 
where $a_{\rm ecl}$ is the Plummer length parameter of stars, 
$M_{\rm gas} = 2 M_{\rm ecl}$, where $M_{\rm ecl}$ is the initial stellar mass of the embedded cluster 
(this implies the star formation efficiency $\sfe \equiv M_{\rm ecl}/(M_{\rm ecl} + M_{\rm gas}) = 1/3$), 
$\tau_{\rm M} = a_{\rm gas}/c_{\rm II} \approx 0.03 \Myr$, 
where $c_{\rm II} = 10 \Kms$ is the sound speed in ionised hydrogen, and
$t_{\rm d} = 0.6 \Myr$ is approximately the lifetime of the ultra-compact HII region and the prior deeply embedded phase.

The simulated clusters orbit a Milky Way-like galaxy on circular trajectories. 
The escaping stars are integrated in the gravitational field of the galaxy, so they form tidal tails, and stellar 
evolution is followed also in the tidal tails. 
The  gravitational potential of the galaxy is represented by the model of \citet{Allen1991}, 
which consists of three components: the bulge modelled as a Plummer potential of mass $1.4 \times 10^{10} \Msun$ and scale-length $0.38 \Kpc$; 
the stellar disc modelled as a Miyamoto-Nagai \citep{Miyamoto1975} potential of mass $8.5 \times 10^{10} \Msun$ and 
radial and vertical scale-lengths $5.3 \Kpc$ and $0.25 \Kpc$, respectively;
and a dark matter halo of a two-component potential which is nearly logarithmic at large galactocentric distances, and which 
encompasses a mass of $8 \times 10^{11} \Msun$ within a sphere of $100 \Kpc$ radius. 
The particular form of the galactic potential has likely negligible influence on the results as we discuss in \refs{ssGalPotential}.

\subsection{Initial conditions} 

\label{ssInitCond}

The clusters are generated by the software package \code{MCLUSTER} \citep{Kupper2011} as Plummer models of stellar mass $M_{\rm ecl}$ and Plummer parameter $a_{\rm ecl}$. 
Employing the relation between the embedded cluster radius and its mass in stars \citep{Marks2012},
\begin{equation}
\frac{a_{\rm ecl}}{\rm{pc}} = 0.077 \left( \frac{M_{\rm ecl}}{\Msun} \right)^{0.13}, 
\label{eMassRadRel}
\end{equation}
we reduce the two parameter space $(M_{\rm ecl}, a_{\rm ecl})$ to one parameter family of models characterised 
by cluster mass $M_{\rm ecl}$. 
The clusters are initially in virial equilibrium with the primordial gas. 
The stars are populated to the cluster according to the recipe of \citet{Aarseth1974b}, 
where the position is generated first from the cluster mass profile, 
whereupon the absolute value of the velocity is drawn from the velocity distribution at the given cluster radius from the Plummer distribution function, 
and the direction of the velocity vector is oriented randomly.
The clusters are initially not mass segregated. 

The initial masses of stars are sampled from the model of the stellar initial mass function (IMF) due to \citet{Kroupa2001a}, where 
the maximum allowed stellar mass, $m_{\rm max}$, within the cluster is a function of $M_{\rm ecl}$ 
according to the $m_{\rm max} - M_{\rm ecl}$ relation \citep{Weidner2010}. 
Thus, only massive clusters harbour the most massive stars. 
The minimum stellar mass is $0.08 \Msun$.
All stars form in binaries with orbital parameters for stars of $< 3 \Msun$ according to the initial distribution function 
for periods, eccentricities and mass ratios of \citet{Kroupa1995a}, 
and more massive stars ($> 3 \Msun$) according to the binary properties of O stars \citep{Sana2012,Moe2017}. 
The method how to generate a realistic initial binary population without affecting the IMF is described by \citet{Oh2015} and \citet{Oh2016}.
Since the evolution of the binaries is at the focus of paper II, the details of the initial binary population and the evolution processes are provided there.

\begin{table*}
\begin{tabular}{ccccccccccc}
Model & $\rm{Nmod}$ & $M_{\rm ecl}$ & $r_{\rm h}$ & $m_{\rm max}$ & $N_{\rm PC}$ & $\sigma(0)$ & $t_{\rm cr}$ & $t_{\rm rlx}$ & $t_{\rm ms}$ & \\
 & & [$\Msun$] & [$\Pc$] & [$\Msun$] & & [$\Kms$] & [$\Myr$] & [$\Myr$] & [$\Myr$] \\
\hline
M1 & 2 &  6400 & 0.299 & 92 & 252 &   10.3  & 0.028 &  58 &  2.10 \\
M2 & 4 &  3200 & 0.274 & 63 & 277 &    7.6  & 0.035 &  40 &  1.44 \\
M3 & 8   &  1600 & 0.251 & 42 & 225 &  5.6  & 0.044 &  28 &  1.00 \\
M4 & 16  &   800 & 0.231 & 26 & 233 &  4.1  & 0.054 &  19 &  0.70 \\
M5 & 32  &   400 & 0.211 & 17 & 214 &  3.1  & 0.068 &  14 &  0.50 \\
M6 & 64  &   200 & 0.194 & 10 & 226 &  2.3  & 0.084 &  10 &  0.37 \\
M7 & 128 &   100 & 0.178 & 6.6 & 217 &  1.7 & 0.104 & 7.7 &  0.28 \\
M8 & 256 &    50 & 0.163 & 4.1 & 151 &  1.2 & 0.129 & 6.1 &  0.22
\end{tabular}
\caption{List of star cluster models.
Model name and number $\rm{Nmod}$ of realisations of the model (cluster of the same global properties 
but with initial conditions from a different random seed) are listed in the first and second columns. 
The following columns list the cluster stellar mass $M_{\rm ecl}$, initial half-mass radius $r_{\rm h}$, 
mass of the most massive star in the cluster $m_{\rm max}$ (mean value from all realisations), 
the number of ProCeps $N_{\rm PC}$ (total for all realisations), 
the initial 3D velocity dispersion $\sigma(0)$, 
the half-mass radius crossing time $t_{\rm cr}$, 
the half-mass radius relaxation time $t_{\rm rlx}$ \citep[][their eqs. 7.106 and 7.108]{Binney2008}, 
and the mass segregation time-scale $t_{\rm ms}$ \citep[][his eq. 9]{Spitzer1969} for $10 \Msun$ stars. 
To calculate the velocity dispersion, $\sigma(0)$, and time-scales $t_{\rm cr}$ and $t_{\rm rlx}$ for embedded clusters, 
we multiply their values for gas-free clusters by $(1/\sfe)^{1/2}$, $(\sfe)^{1/2}$ and $(1/\sfe)^{3/2}$, respectively.
}
\label{tsimList}
\end{table*}

\subsection{Assumptions about the embedded cluster mass function}

\label{ssECMF}

There is observational evidence that the embedded cluster mass function (ECMF)
in the Galaxy \citep[e.g.][]{Lada2003,FuenteMarcos2004} 
as well as in more distant galaxies \citep[e.g.][]{Whitmore1999,Bik2003,Gieles2006} can 
be approximated by a power law of slope $-2$, i.e.
\begin{equation}
\xi_{\rm ecl}(M_{\rm ecl}) \equiv \nder{N_{\rm ecl}(M_{\rm ecl})}{M_{\rm ecl}} \propto M_{\rm ecl}^{-2},
\label{eClusterIMF}
\end{equation}
where $\dd N_{\rm ecl}(M_{\rm ecl})$ is the number of embedded star clusters forming in a mass bin of size $\dd M_{\rm ecl}$.

Our models span cluster mass range of $M_{\rm ecl} \in (50 \Msun, 6400 \Msun)$, where the lower mass 
limit follows from the $m_{\rm max} - M_{\rm ecl}$ relation, which requires $m_{\rm max} \gtrsim m_{\rm min,Ceph}$, and 
the upper mass limit is comparable to the mass of the most massive young star clusters observed in the galaxy M~31 
($\approx 8.5 \times 10^3 \Msun$; \citealt{Johnson2017}).
The dependence $\xi_{\rm ecl}(M_{\rm ecl}) \propto M_{\rm ecl}^{-2}$ has the convenient property 
that the total mass of clusters in each logarithmic mass bin of constant size contains an identical total mass of clusters regardless of $M_{\rm ecl}$.


\subsection{The library of simulations}

\label{ssLibrary}


The cluster masses, $M_{\rm ecl}$, span the range of $50 \Msun$ to $6400 \Msun$, and they are 
sampled by multiples of $2$, i.e. $M_{\rm ecl} = 50 \Msun, 100 \Msun, .., 6400 \Msun$. 
The models are listed in \reft{tsimList}. 
We run each model Nmod times with different random seed for stellar masses, positions and velocities, 
where Nmod$=2^{i}$, and $i$ is the index after the model name, e.g. 
models M4 have $\rm{Nmod}=16$.
We choose these values of Nmod to have the same total mass of star clusters in each logarithmic cluster mass bin $\dd M_{\rm ecl}$ 
when used with the ECMF of \eq{eClusterIMF}.
This facilitates calculating any quantity $\langle q \rangle$ for the whole population of clusters (and thus its observed value integrated over a galaxy) as 
$\langle q \rangle = (\sum_{i=1}^8 q_i w_{i})/(\sum_{i=1}^8 w_{i})$, where $q_i$ is the quantity of interest and $w_{i}$ 
the relevant weighting factor (e.g. number of all Cepheids originating in all clusters of index $i$
\footnote{Although the total mass of clusters in each logarithmic mass bin is the same, the number of stars which evolve to Cepheids 
is different in each bin because of the $m_{\rm max} - M_{\rm ecl}$ 
relation. For example, the least massive clusters do not populate the stellar IMF up to the maximum mass $m_{\rm max,Ceph}$ of stars which evolve to Cepheids 
because they have $m_{\rm max} < m_{\rm max,Ceph}$.}
). 

We calculate the population of clusters described in the previous paragraph for three galactocentric 
radii: $R_{\rm g} = 4 \Kpc$, $R_{\rm g} = 8 \Kpc$, and $R_{\rm g} = 12 \Kpc$ for solar metallicity, i.e. $Z = 0.014$. 
We also calculate the same population of clusters for galactocentric radius $R_{\rm g} = 8 \Kpc$ and 
subsolar metallicity of $Z = 0.006$ and $Z = 0.002$. 
Thus, for each considered choice of $R_{\rm g}$ and $Z$, we calculate 510 simulations in total.

\section{Identification of Cepheids}

\label{sIdentification}



Isolated stars with zero age main sequence (ZAMS) mass in the appropriate mass 
range $(m_{\rm min,Ceph},m_{\rm max,Ceph})$ enter 
the Cepheid instability strip in the course of their evolution. 
Present models take into account the metallicity dependence of both the stellar evolutionary tracks \citep{Hurley2000} and 
the borders of the instability strip for Cepheid stars \citep{Anderson2016}. 
They provide the mass range for Cepheids as follows:
$(m_{\rm min,Ceph},m_{\rm max,Ceph}) = (3.2 \Msun, 11.0 \Msun)$ for $Z = 0.002$, 
$(m_{\rm min,Ceph},m_{\rm max,Ceph}) = (4.0 \Msun, 12.0 \Msun)$ for $Z = 0.006$, and 
$(m_{\rm min,Ceph},m_{\rm max,Ceph}) = (4.7 \Msun, 13.0 \Msun)$ for $Z = 0.014$.
For the purpose of this work, we term all stars with initial 
mass in the range of $(m_{\rm min,Ceph},m_{\rm max,Ceph})$ prospective Cepheids (ProCephs), 
and stars with initial mass below $m_{\rm min,Ceph}$ as lower mass stars. 
The star is then referred to as ProCep or a lower mass star for the rest of the simulation regardless of its actual mass, which can change with time. 
If the ProCep becomes a Cepheid, it is denoted a canonical Cepheid (CanCeph). 
There are also Cepheids which originate from a coalescence in a binary with both components having a ZAMS mass below $m_{\rm min,Ceph}$; 
we call these Cepheids as upscattered Cepheids (UpCeps).
For the adopted evolutionary tracks, stars with mass $m_{\rm min,Ceph}$ terminate core Helium burning at time $t_{\rm He,end}$ of 
$300 \Myr$, $190 \Myr$ and $146 \Myr$ for $Z = 0.002$, $Z = 0.006$, and $Z = 0.014$, respectively. 

\iffigscl
\begin{figure*}
\includegraphics[width=\textwidth]{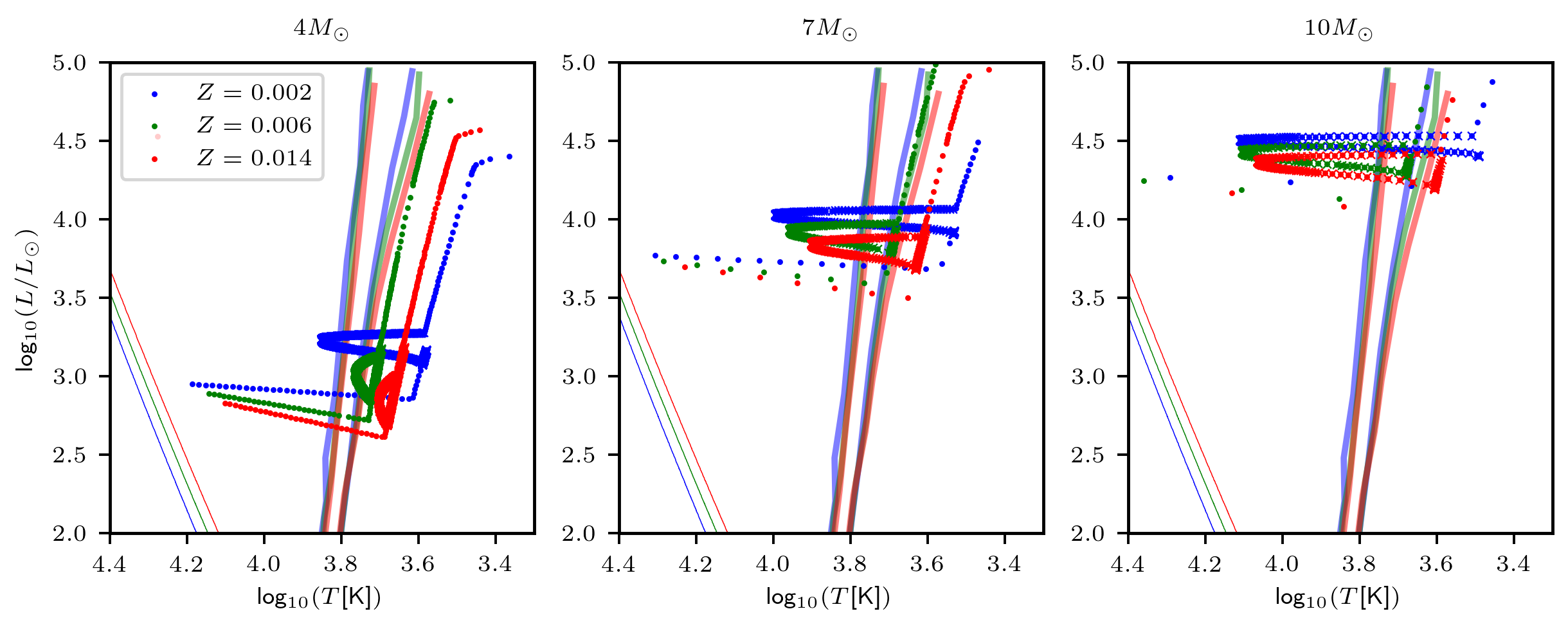}
\caption{HR diagrams outputted from \nbdvi for single stars of ZAMS mass of 
$4 \Msun$ (left panel), $7 \Msun$ (middle panel), and $10 \Msun$ (right panel). 
The evolutionary paths are plotted after the stars left the main sequence.
The colour indicates the metallicity, which is $Z = 0.002$ (blue), $Z = 0.006$ (green) and $Z = 0.014$ (red).
Large crosses denote the time of core helium burning.
For the purpose of this plot, we selected single stars, which evolve well in isolation from the other stars in the cluster.
Thick solid lines show the Cepheid instability strip for the metallicities considered \citep{Anderson2016}.
Thin solid lines represent ZAMS for stars of corresponding metallicity.}
\label{fHrexample}
\end{figure*} \else \fi


We search Cepheids automatically in the HR diagrams produced by \nbdvid. 
For each star in the simulation, we record its temperature and luminosity whenever its stellar evolution is updated, 
which, during core helium burning, typically occurs every $0.01$ to $0.3 \Myr$ depending on the Cepheid mass and exact position in the HR diagram. 
In this way, we obtain densely sampled HR diagrams even during the rapid evolution on the instability strip as shown in \reff{fHrexample}. 
Then, the algorithm diagnoses each star and classifies it as a Cepheid if the star fulfils all of the following criteria: 
(i) The star's initial mass is in the mass range $(m_{\rm min,Ceph}/2, m_{\rm max,Ceph})$, the lower value of the mass range is to include 
a possible merger of two $0.5 m_{\rm min,Ceph}$ stars in a hard binary, which can evolve to a Cepheid (UpCeps). 
%
%
(ii) The star undergoes core Helium burning.  
(iii) The surface temperature at which core Helium burning starts is rightwards from 
the blue boundary of the instability strip to exclude the stars which do not have Cepheid loops (stars with 
$m > m_{\rm max,Ceph}$ typically have a monotonically decreasing temperature during core Helium burning). 
(iv) The maximum temperature during core helium burning is leftwards from the red boundary of the 
instability strip, i.e. the star lies on the instability strip during its core Helium burning at least on one output file. 
We use metallicity-dependent boundaries of the instability strip from \citet{Anderson2016}.
Verification of this searching procedure for more complicated HR diagrams  
with merging stars is presented in Appendix \ref{saMergers}. 

To obtain the properties of the Cepheid (e.g. its distance from the cluster centre or binarity), we first determine the times when 
the Cepheid crosses the instability strip and then evaluate the desired properties at 
a random time (the event when the Cepheid is observed) from the time interval when the star is located within the instability strip. 
If the Cepheid passes the instability strip twice (we ignore the first passage through the 
instability strip because of its very short duration), we determine the time 
randomly from one of the two intervals on the instability strip weighted by their duration. 
The dynamics of Cepheids usually changes little when being within the strip because the duration in the strip is significantly shorter than the 
cluster relaxation time. 

We classify the Cepheid as a cluster Cepheid if it is located closer than $10 \Pc$ to the density centre \citep{Casertano1985} of the cluster.  
The majority of Cepheids bound to the cluster are located
near the cluster centre (closer than $\approx 2 \Pc$) as a result of the cluster half-mass radii (several parsec) and mass segregation of ProCeps.
The distance of $10 \Pc$ is comparable to the tidal radius of the clusters, so if a star exceeds this distance, it is likely to become unbound and recedes. 
If the cluster is already dissolved, all its Cepheids are classified as field Cepheids. 

To measure the fraction of Cepheids located in clusters (relatively to the Cepheids in the field and in clusters combined) 
in different contexts, we introduce the following terminology. 
The fraction of Cepheids in clusters of initial mass $M_{\rm ecl}$ at given time $t$ is denoted $\pccii$.
The fraction of Cepheids in clusters of a particular cluster mass $M_{\rm ecl}$ throughout the time interval $(0, 300 \Myr)$ 
is denoted $\pcci$, 
\begin{equation}
\pcci = \frac{\sum_i \; p_{\rm CC} (M_{\rm ecl}, t_i) \Delta t_i}{\sum_i \; \Delta t_i},
\label{epcci}
\end{equation}
where $\Delta t_i$ is the time interval near time $t_i$, and the summation goes over all time intervals. 
The galaxy wide fraction of Cepheids in clusters, i.e. formed in all star clusters within the galaxy over their 
life-time is denoted $f_{\rm CC}$, 
\begin{equation}
f_{\rm CC} = \frac{\sum_i \; \xi_{\rm ecl}(M_{{\rm ecl},i}) p_{\rm CC} (M_{{\rm ecl},i}) \Delta M_{{\rm ecl},i}}{\sum_i \; \xi_{\rm ecl}(M_{{\rm ecl},i}) \Delta M_{{\rm ecl},i}}, 
\label{efcc}
\end{equation}
where $\xi_{\rm ecl}(M_{\rm ecl})$ is the ECMF of \eq{eClusterIMF}, and the summation goes over all cluster mass intervals.
We stress that throughout this work, $M_{\rm ecl}$ denotes the initial (not actual) cluster mass. 
As the cluster evolves, its actual mass can be significantly lower than its initial value. 
In addition, we study the fraction of Cepheids in clusters $\pcciii$ which are of a given stellar mass $m_{\rm Ceph}$ regardless of the mass of 
their birth star clusters.

To gain basic insight, it is useful to mention some typical numbers of ProCeps formed in a coeval stellar population (CSP).
Consider a CSP, which is populated by the Kroupa IMF from the mass interval of $0.08 \Msun$ to $50 \Msun$. 
For such a CSP, the total mass of ProCeps 
comprises $17$ \%, $14$ \% and $13$ \% of the total initial CSP mass 
for $Z = 0.002$, $Z = 0.006$ and $Z = 0.014$, respectively. 
By number, ProCeps constitute $1.7$ \%, $1.2$ \% and $0.95$ \% of the total initial CSP 
for $Z = 0.002$, $Z = 0.006$ and $Z = 0.014$, respectively.

\section{Cepheids in dynamically evolving star clusters} 

\label{sDynamical}

In order to quantify the position of Cepheids relative to their birth clusters, 
we study the influence of the dynamical evolution of star clusters on ProCeps. 
First, we review the most relevant 
features of the dynamics of young star clusters experiencing expulsion of their residual gas (\refs{ssOverviewDyn}). 
Then, we investigate the mechanisms which unbind ProCeps from 
clusters and therefore determine the value of $f_{\rm CC}$ as a function of cluster age and 
mass for clusters orbiting their host galaxy at a radius of $R_{\rm g} = 8 \Kpc$ and of metallicity $Z = 0.014$ (\refs{ssCephDyn}). 
We discuss the dependence of $f_{\rm CC}$ on $R_{\rm g}$ and $Z$ in Sections \ref{ssGalRad} and \ref{ssZmet}. 
The probability to find a star cluster hosting a particular number of Cepheids at the same time is studied in \refs{ssMultCep}, 
and the Cepheids in the field, but still close to their birth star clusters are dealt with in \refs{ssOutskirts}.


\subsection{Overview of dynamical evolution of tidally-limited star clusters experiencing early gas expulsion}

\label{ssOverviewDyn}


\iffigscl
\begin{figure}
\includegraphics[width=\columnwidth]{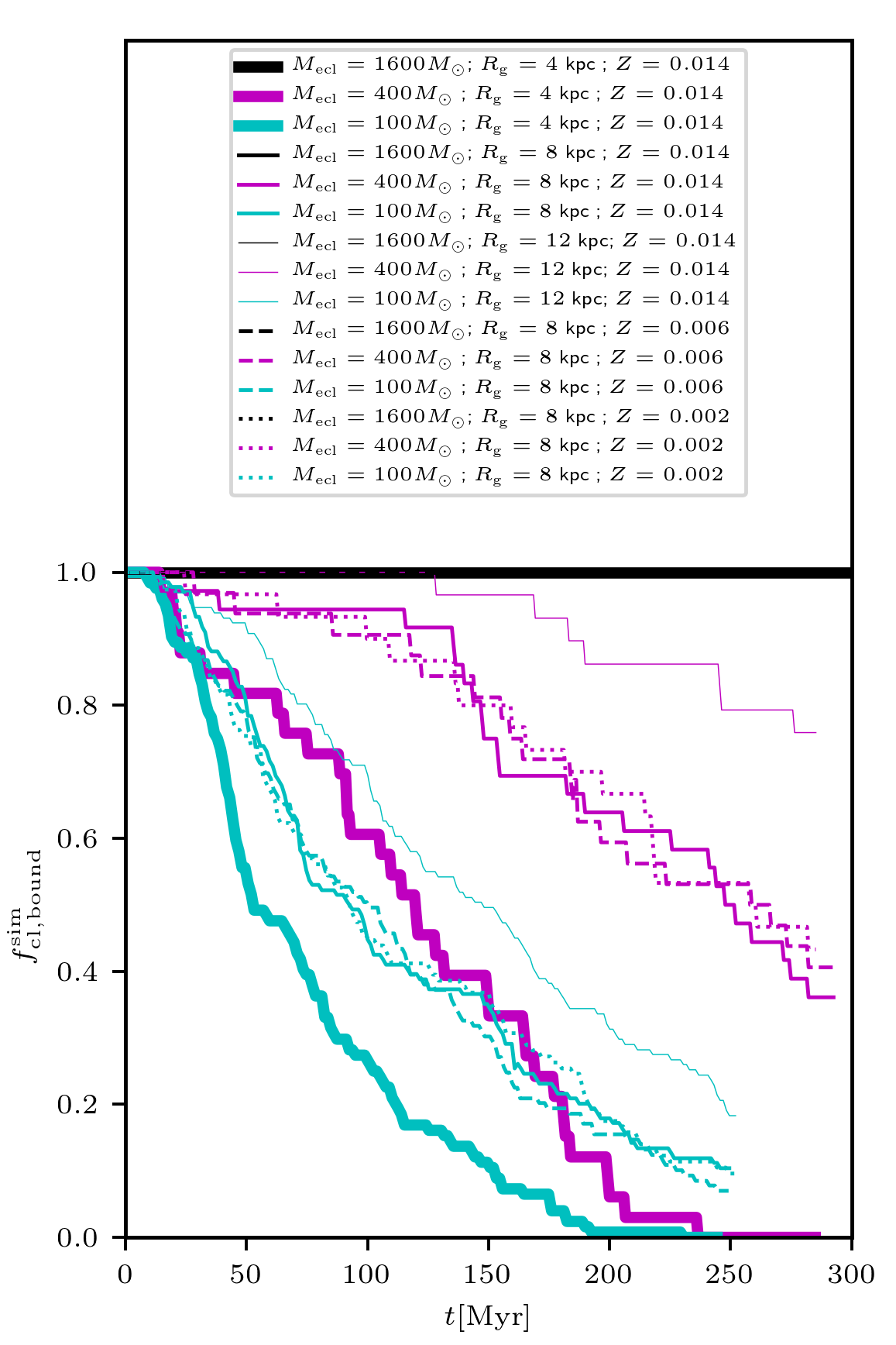}
\caption{The fraction of star clusters 
$f_{\rm cl,bound}^{\rm sim}$ which are gravitationally bound by time $t$ in the external tidal field of a galaxy as a function of the 
galactocentric radius $R_{\rm g}$, initial cluster mass $M_{\rm ecl}$ and metallicity $Z$. 
The galactocentric radius $R_{\rm g}$ of the orbit is indicated by 
the line thickness (thick - $4 \Kpc$; medium - $8 \Kpc$; thin - $12 \Kpc$), 
the cluster mass is indicated by the line colour (cyan - $100 \Msun$; magenta - $400 \Msun$; black - $1600 \Msun$), 
and the metallicity for the models with $R_{\rm g} = 8 \Kpc$ is indicated by the line style (solid - $Z = 0.014$; dashed - $Z = 0.006$; dotted - $Z = 0.002$).}
\label{fdissolved}
\end{figure} \else \fi

In the present models, stars do not conserve their adiabatic 
invariants in the gravitational potential during gas expulsion because the 
gas expulsion time scale $\tau_{\rm M}$ is substantially shorter ($\tau_{\rm M} \approx 0.02 \Myr$)
than the stellar half-mass crossing time, $t_{\rm cr}$ (see \reft{tsimList}).
As a consequence, the cluster substantially expands while most stars ($\approx 70$ \% of the initial 
cluster population) escape from the cluster \citep{Tutukov1978,Lada1984,Geyer2001,Kroupa2001b,Baumgardt2007}. 

On a time-scale of $1 \Myr$ to several tens of Myr (depending on the cluster mass), 
the cluster revirialises, settling at a significantly larger half-mass radius than was its initial value \citep{Banerjee2013,Banerjee2017}. 
Further evolution is driven purely by stellar dynamics, where stars 
escape from the cluster mainly due to slow evaporation with a few energetic ejections, which occur mostly between more massive stars 
\citep{Baumgardt2003,Fujii2011,Oh2015,Schoettler2020}. 
As the cluster mass decreases due to escaping stars and stellar mass-loss, the tidal radius decreases,
thereby further facilitating the escape of stars until the cluster eventually 
dissolves completely, leaving behind an unbound stellar stream.

Figure \ref{fdissolved} shows the fraction of star clusters that remain gravitationally bound\footnote{In this work, we consider a star cluster to be any group of at least 
ten c.m. bodies (a c.m. body is either a single star or a regularised binary or hierarchical system), which is located within the cluster tidal radius. 
}
at a given age $t$ in the tidal field of a galaxy (taking here the Galaxy as the specifically computed example). 
At a particular galactocentric radius, the dissolution time of a cluster is mainly determined by the cluster mass $M_{\rm ecl}$. 
Particularly, at $R_{\rm g} = 8 \Kpc$, all clusters with $\gtrsim 1600 \Msun$ survive up to the age of $300 \Myr$ (the black lines in \reff{fdissolved}), 
while half of all $400 \Msun$ and $100 \Msun$ clusters dissolve within $250 \Myr$ (medium magenta line) and $90 \Myr$ (medium cyan line), respectively. 

Another important parameter strongly impacting the cluster dissolution time-scale is the galactocentric radius of its orbit \citep{Vesperini1997,Baumgardt2003}. 
Clusters orbiting closer to the galactic centre dissolve earlier due to the stronger tidal field:  
at $R_{\rm g} = 4 \Kpc$ (thick solid lines in \reff{fdissolved}), 
half of $400 \Msun$ and $100 \Msun$ clusters dissolve within $120 \Myr$ and 
$50 \Myr$, respectively.
At $R_{\rm g} = 12 \Kpc$ (thin solid lines), the majority ($\approx$ 75 \%) of $400 \Msun$ clusters survive until at least $300 \Myr$, 
while half of $100 \Msun$ clusters dissolve within $150 \Myr$. 
All $1600 \Msun$ clusters are preserved at least for $300 \Myr$ even at the smallest radius of $4 \Kpc$. 
The influence of metallicity on cluster dissolution is negligible as shown by the dashed and dotted lines for the model with $R_{\rm g} = 8 \Kpc$.

\iffigscl
\begin{figure}
\includegraphics[width=\columnwidth]{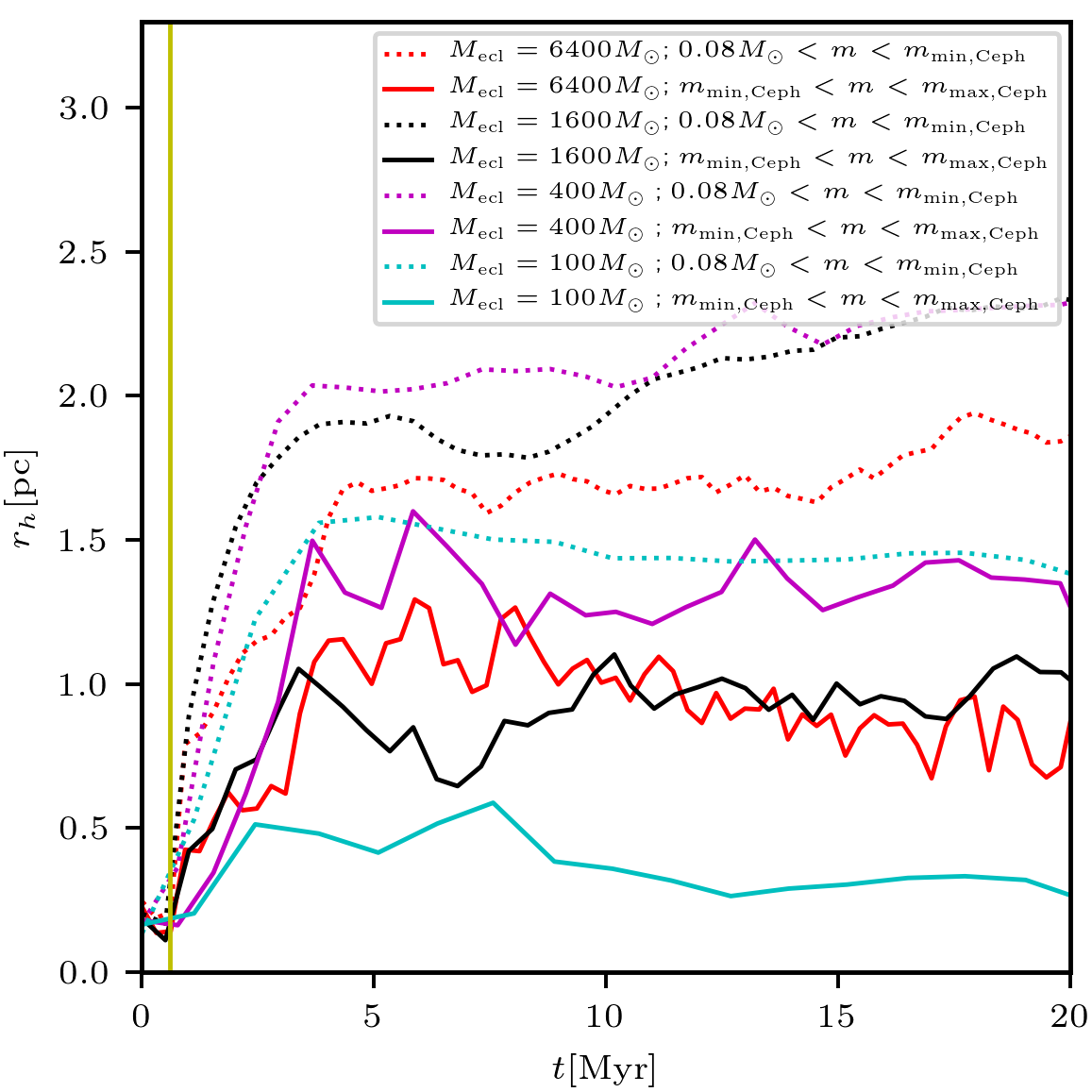}
\caption{Evolution of the half-mass radius $r_{\rm h}$ of ProCeps (solid lines) and lower mass stars (dotted lines) for 
clusters of mass $M_{\rm ecl} = 100 \Msun$, $400 \Msun$, $1600 \Msun$ and $6400 \Msun$. 
The clusters orbit the galaxy at $R_{\rm g} = 8 \Kpc$ and have metallicity $Z = 0.014$.
The vertical yellow line shows the time of gas expulsion $t_{\rm d}$. 
The least massive clusters ($M_{\rm ecl} = 100 \Msun$; cyan) mass segregate their ProCeps before $t_{\rm d}$,  which results 
in ProCeps being more concentrated than lower mass stars. 
The following gas expulsion then preferentially unbinds stars from outskirts of the cluster, which are lower mass stars, resulting 
in a cluster which contains disproportionately more ProCeps. 
In contrast, the more massive clusters do not mass segregate before $t_{\rm d}$, 
so the stellar loss due to gas expulsion affects ProCeps in a similar way as lower mass stars.}
\label{fmassSeg}
\end{figure} \else \fi

\subsection{Dynamical evolution of prospective Cepheids in clusters at $R_{\rm g} = 8 \Kpc$ and $Z = 0.014$}

\label{ssCephDyn}

Here, we investigate the significance of different physical mechanisms which release ProCeps from clusters. 
Stars generally escape their birth clusters by three mechanisms: the rapid change of the gravitational potential 
during gas expulsion; gradual evaporation due to encounters within the cluster; and 
fast ejections in close encounters between several bodies. 

To identify the mechanism responsible for the escape of the star, 
we define that any star escaping at speed $v_{\rm esc} > 3 \sigma(0)$, where $\sigma(0)$ is the initial 
3D velocity dispersion of stars in the embedded cluster (\reft{tsimList}), is an ejection. 
It is unlikely that a star achieves such a high speed without interacting in an unstable system of three or more bodies. 
The cause of an escape of a star with $v_{\rm esc} < 3 \sigma(0)$ is identified as 
the gas expulsion if the star escapes at time $t < t_{\rm ee}$, where $t_{\rm ee} = r_{\rm t}/(0.1 \sigma(0))$ and $r_{\rm t}$ 
is the tidal radius of the cluster; or as dynamical evaporation if the star escapes at time $t > t_{\rm ee}$.
The choice of $t_{\rm ee}$ is motivated by the typical value of the velocity of stars released due to gas expulsion, which is of the order of $\sigma(0)$. 
These stars reach the tidal boundary of the cluster at time $\approx r_{\rm t}/\sigma(0)$. 
The factor of $0.1$ is a safety factor to take into account the low velocity tail of the velocity distribution. 
The same definition of $t_{\rm ee}$ was adopted by \citet{Dinnbier2020a}.

\iffigscl
\begin{figure}
\includegraphics[width=\columnwidth]{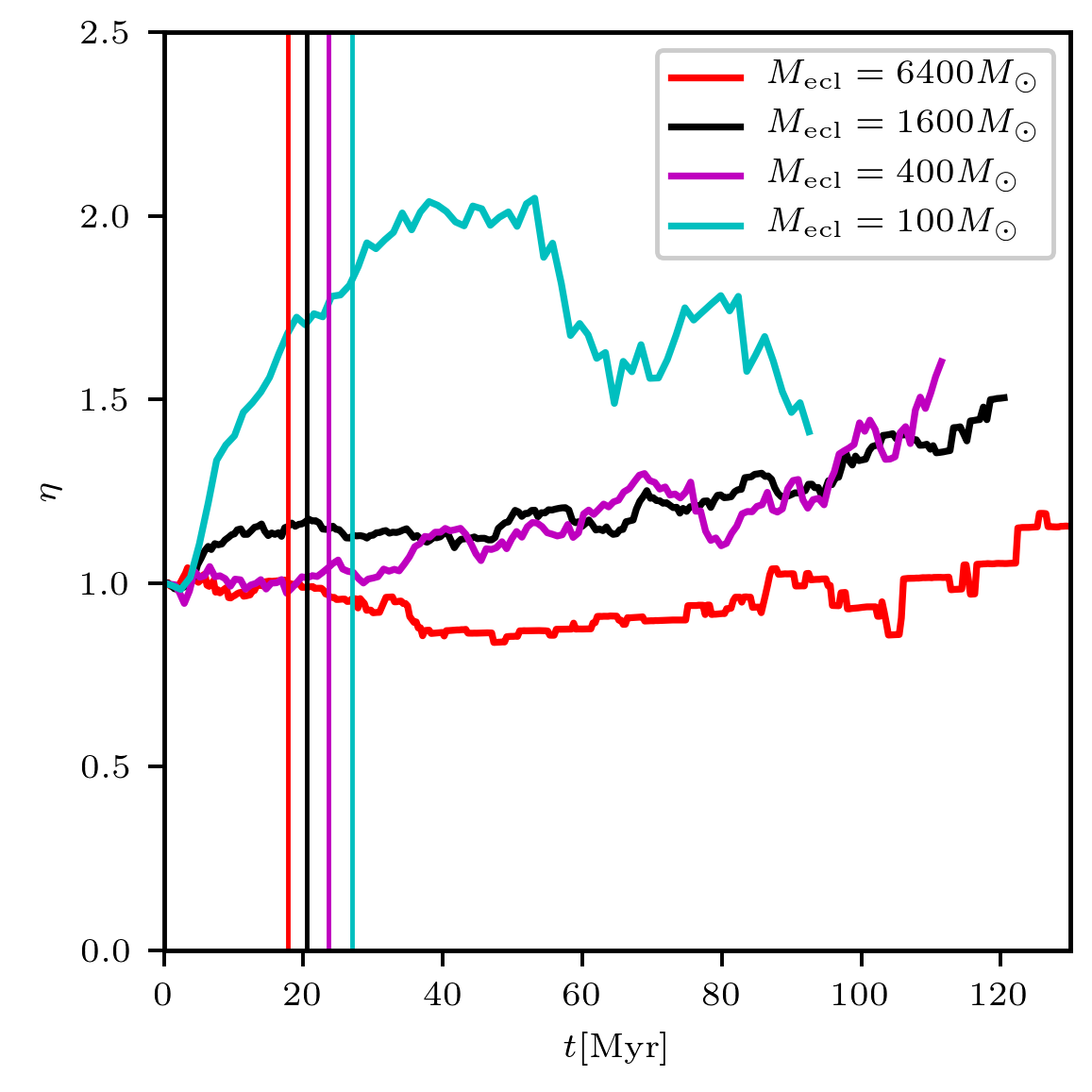}
\caption{
The abundance of ProCeps $\eta$ relative to the lower mass stars in clusters as a function of time $t$.
The time $t_{\rm ee}$, indicating the transition between stars escaping due to gas expulsion ($t < t_{\rm ee}$) and
evaporation ($t > t_{\rm ee}$) is indicated by vertical lines.
While the non-mass segregated clusters ($M_{\rm ecl} \gtrsim 400 \Msun$) have only small changes in $\eta$, 
the mass segregated cluster ($M_{\rm ecl} = 100 \Msun$) has an increase in $\eta$ around time $t_{\rm ee}$. 
To avoid rapid fluctuations due to small number statistics, the plots are terminated when there is less than $10$ ProCeps in clusters in all 
realisations of clusters of given mass.
The figure represents clusters at $R_{\rm g} = 8 \Kpc$ and $Z = 0.014$. 
}
\label{fnCeph1}
\end{figure} \else \fi

\iffigscl
\begin{figure}
\includegraphics[width=\columnwidth]{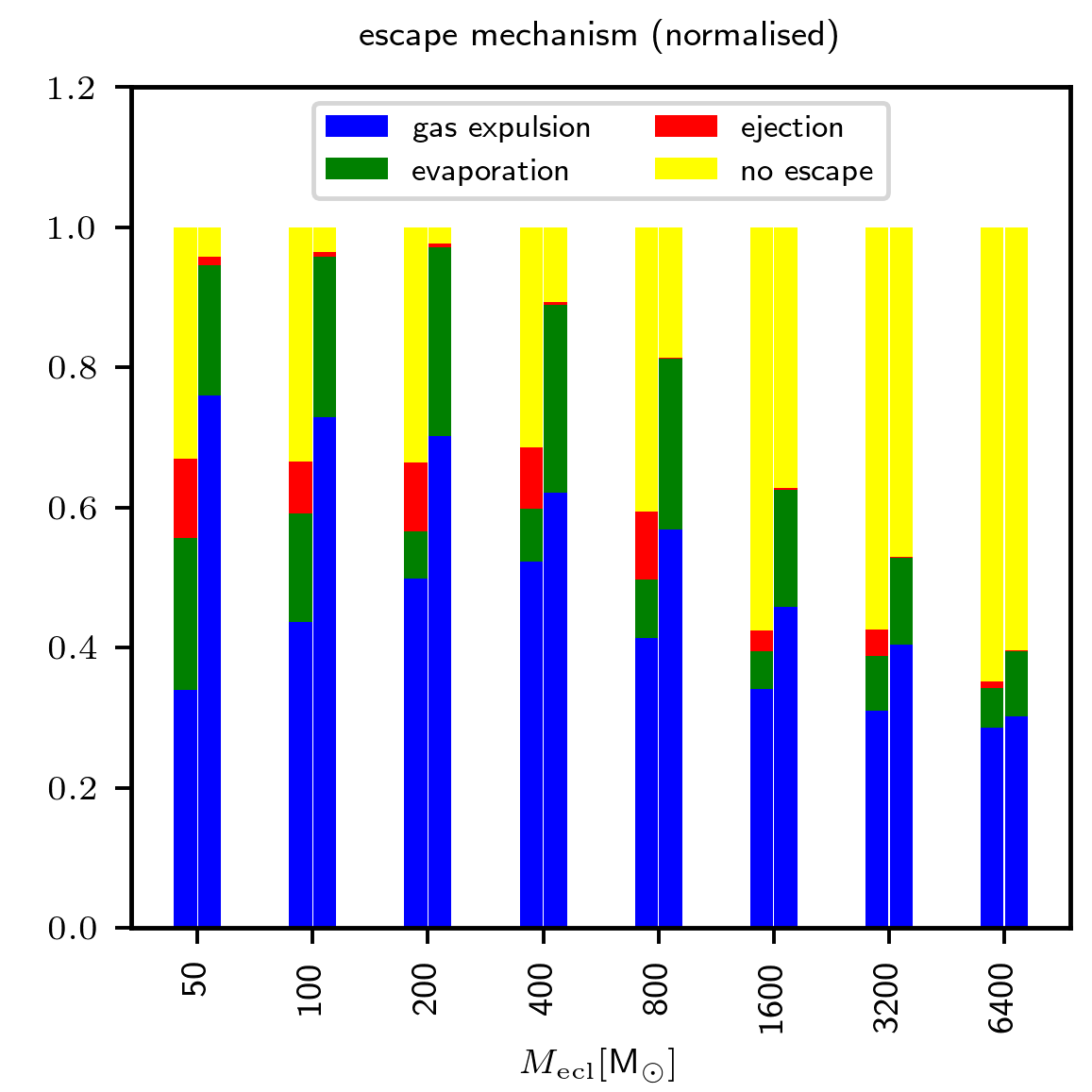}
\caption{
The dynamical mechanisms responsible for stellar escapes as a function of the cluster mass.
For each cluster mass, we plot two bar plots; the left bar plot represents ProCeps,
the right bar plot represents lower mass stars.
The colours indicate the relative number of stars in each group which escaped
due to gas expulsion (blue), dynamical evaporation (green), dynamical ejection (red), and the rest which
either terminated their core Helium burning within the cluster (for ProCeps) or are still present in the cluster by the age of $145 \Myr$ (yellow).
Note that in the least massive clusters ($M_{\rm ecl} \lesssim 200 \Msun$), which mass segregate before $t_{\rm d}$,
lower mass stars and ProCeps react differently on gas expulsion:
as $M_{\rm ecl}$ decreases, more low mass stars get unbound, while fewer ProCeps get unbound.
The figure is for clusters with $R_{\rm g} = 8 \Kpc$ and $Z = 0.014$.
}
\label{fnCeph2}
\end{figure} \else \fi

The early evolution of embedded star clusters is dominated by gas expulsion, which in our models starts at $t_{\rm d} = 0.6 \Myr$ 
after formation of the cluster. 
However, the clusters dynamically evolve prior to $t_{\rm d}$. 
We compare mass segregation of ProCeps and of lower mass stars by measuring the half-mass radius, $r_{\rm h}$, 
for both stellar populations, which is shown in \reff{fmassSeg}. 
The simulation of the $M_{\rm ecl} = 100 \Msun$ cluster has $r_{\rm h}$ for ProCeps significantly smaller than $r_{\rm h}$ for lower mass stars 
at $\approx t_{\rm d}$, which means that ProCeps are concentrated near the cluster centre already at the time of gas expulsion. 
The preferential presence of ProCeps near the cluster centre 
makes them more resilient to gas expulsion, which unbinds stars mainly from the outskirts of the cluster. 
These are mainly lower mass stars that are located at the outskirts. 
After gas expulsion, this cluster contains an excess of ProCeps relatively to lower mass stars. 
More massive clusters ($M_{\rm ecl} \gtrsim 400 \Msun$) do not have enough time to mass segregate prior $t_{\rm d}$, 
which means that gas expulsion affects ProCeps and lower mass stars to a similar extent, meaning that more massive clusters lose 
ProCeps at a similar rate as lower mass stars. 
The mass segregation time-scale for ProCeps seen in the simulations is close to the analytic value $t_{\rm ms}$ 
obtained from \citet{Spitzer1969}, which is listed in \reft{tsimList}; 
for $M_{\rm ecl} = 100 \Msun$, the simulated cluster 
mass segregates before $t_{\rm d}$  (\reft{tsimList} provides $t_{\rm ms} = 0.28 \Myr$). 
For $M_{\rm ecl} \gtrsim 400 \Msun$, the simulated cluster mass segregates around or after $t_{\rm d}$ (\reft{tsimList} provides $t_{\rm ms} \gtrsim 0.50 \Myr$)
\footnote{The formula used for calculating $t_{\rm ms}$ according to \citet{Spitzer1969} was derived under idealised assumptions, so it might vary by a factor of several.
The close agreement between our simulations is probably coincidental for our particular cluster profile, IMF and the choice of stars with the 
mass $> 10 \Msun$ to be mass segregated.}
.
%
%
Mass segregation continues after revirialisation, but it occurs on a significantly longer time-scale 
($t_{\rm ms} \approx 30 \Myr$) than its initial value $t_{\rm ms}$, because gas expulsion decreased 
the density of the cluster by a factor of $10^3$.  


To which extent does the preferential escape of lower mass stars as a consequence of mass segregation increase the relative population 
of ProCeps, and make the lower mass clusters "Cepheid rich"? 
To answer this question, we compare the fraction of ProCeps in clusters (i.e. $N_{\rm PC,cl}/N_{\rm PC,tot}$) to the fraction of 
lower mass stars in clusters (i.e. $N_{\rm lm,cl}/N_{\rm lm,tot}$), and define 
the relative abundance of ProCeps as
\begin{equation}
\eta \equiv \frac{N_{\rm PC,cl}/N_{\rm PC,tot}}{N_{\rm lm,cl}/N_{\rm lm,tot}}, 
\label{eOverAbund}
\end{equation}
and plot it as a function of time (\reff{fnCeph1}). 
If $\eta$ is larger than $1$ (lower than $1$), 
the cluster has an excess (deficit) in ProCeps relative to the lower mass stars and thus to the overall stellar content because 
the lower mass stars dominate the overall population by number.
At time $t_{\rm ee}$, $\eta$ is close to $1$ for clusters with $M_{\rm ecl} \gtrsim 400 \Msun$, 
but it is significantly larger ($\eta \approx 1.8$) for the $100 \Msun$ clusters which mass segregated before gas expulsion. 
After $50 \Myr$, $\eta$ decreases in these clusters because of two effects. 
First, close encounters between ProCeps in the cluster centre eject ProCeps more likely than lower mass stars (see \reff{fnCeph2} below). 
Second, mass segregation retains more massive ProCeps in clusters, while the less massive ProCeps are more likely to escape to the field. 
Since more massive ProCeps evolve to Cepheids earlier than the less massive ones, ProCeps in clusters are preferentially depleted as a result of  
stellar evolution only.
Thus, lower mass clusters contain more ProCeps than what would be expected from their total mass. 

Clusters in the mass range $\approx 400 \Msun$ to $\approx 1600 \Msun$ gradually increase their $\eta$ with time as the 
rarefied post gas expulsion clusters mass segregate (\reff{fnCeph1}). 
The most massive clusters in our study ($M_{\rm ecl} = 6400 \Msun$) have $\eta$ close to $1$ almost independent of their age because 
of their very long mass segregation time-scale.  
The differences in dynamical evolution of ProCeps between a lower mass ($100 \Msun$; movie 1) and more massive ($3200 \Msun$; movie 2) star cluster are 
illustrated in \reff{fLowHighMassCluster} of Appendix \ref{saLowHighMassCluster}, and in the accompanying online material
\footnote{https://zenodo.org/record/5786512}
.


\iffigscl
\begin{figure}
\includegraphics[width=\columnwidth]{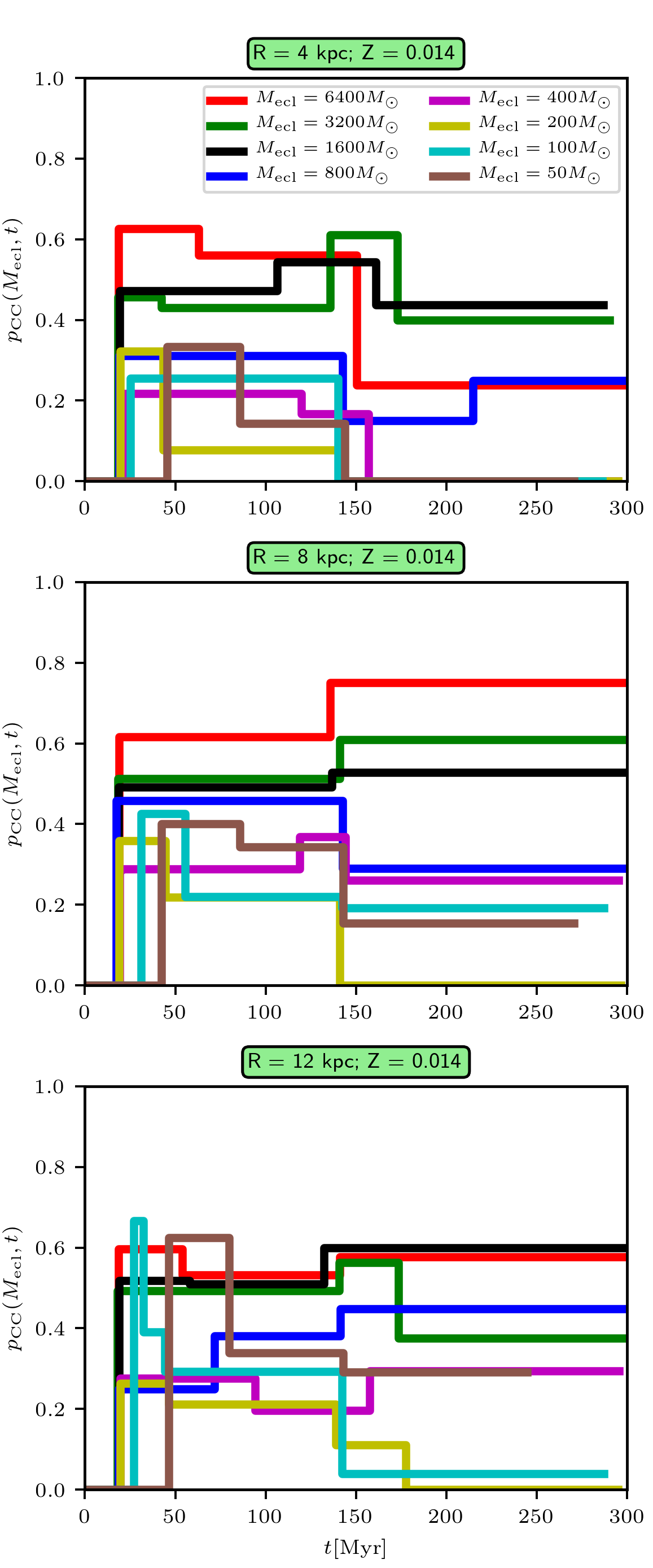}
\caption{
The fraction $\pccii$ of cluster Cepheids as a function of 
time for clusters orbiting the galaxy at different galactocentric radii: 
$R_{\rm g} = 4 \Kpc$ (upper panel), $R_{\rm g} = 8 \Kpc$ (middle panel), and $R_{\rm g} = 12 \Kpc$ (lower panel). 
The cluster mass is indicated by colour. 
Metallicity is $Z = 0.014$ for all models. 
The plot shows all Cepheids including UpCeps, which occur after $t_{\rm He,end}$.
Generally, $\pccii$ decreases with time for lower mass clusters ($M_{\rm ecl} \lesssim 800 \Msun$), and it stays approximately constant 
for clusters more massive than that.
The histogram is binned by the Bayesian Blocks algorithm. 
The lines terminate at $300 \Myr$ or at the occurrence of the last Cepheid.
}
\label{fcluster_inOutRad}
\end{figure} \else \fi

\iffigscl
\begin{figure*}
\includegraphics[width=\textwidth]{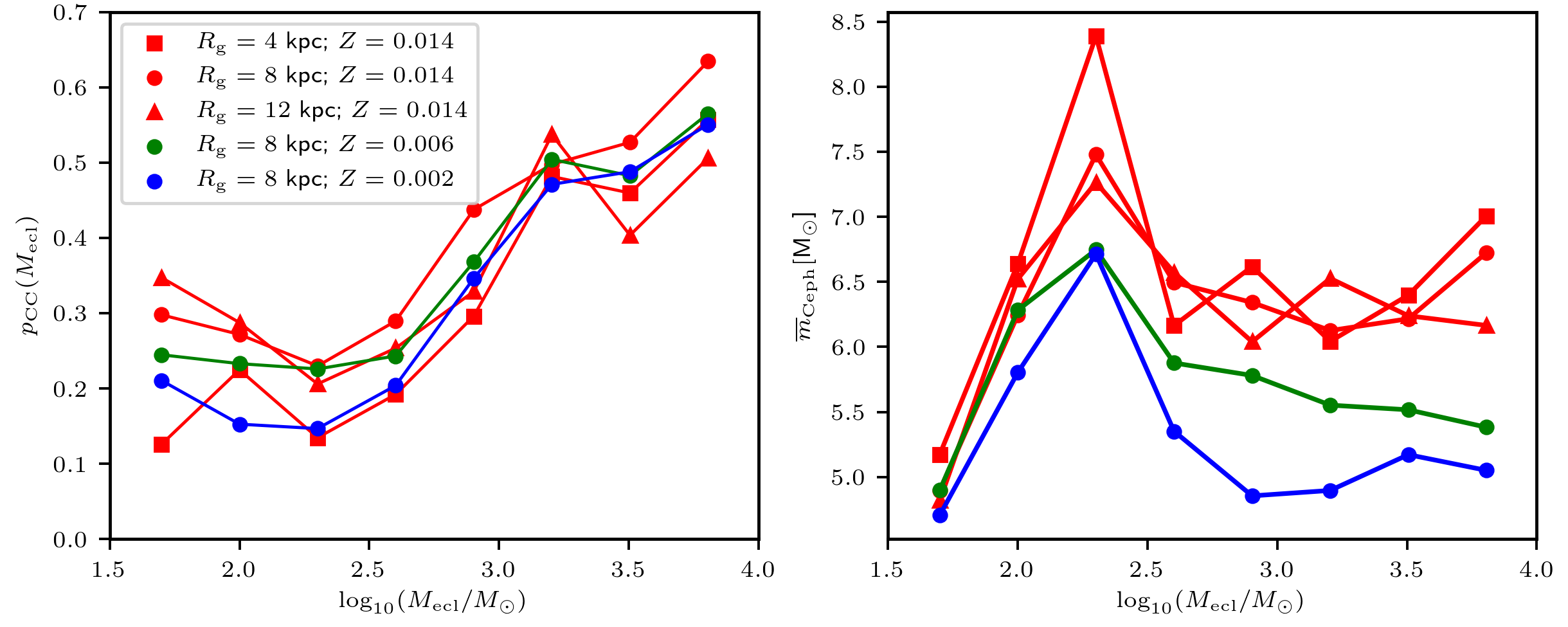}
\caption{\figpan{Left panel:} The fraction of cluster Cepheids $\pcci$ (including UpCeps) as a function of cluster mass $M_{\rm ecl}$ integrated over 
the first $300 \Myr$ of the cluster evolution. 
Clusters orbiting at different galactocentric radii $R_{\rm g}$ are denoted by different symbols; 
clusters with different metallicity $Z$ are denoted by different colour of the symbol. 
The external field of the galaxy, which impacts more the lower mass clusters, is responsible for the decrease of $\pcci$ with $R_{\rm g}$. 
$\pcci$ increases with $M_{\rm ecl}$ because more massive clusters withstand better gas expulsion and the tidal field of the galaxy. 
$\pcci$ for lower mass clusters decreases with metallicity because Cepheids occur later at lower $Z$, at which time more low mass clusters has been 
dissolved. 
\figpan{Right panel:} The mean mass of Cepheids in the cluster as a function of $M_{\rm ecl}$. 
At a given metallicity, $\overline{m}_{\rm Ceph}$ peaks at a cluster mass of $200 \Msun$ because lower mass clusters do not live 
long enough to host lower mass Cepheids.
The decrease of $\overline{m}_{\rm Ceph}$ with decreasing $Z$ is caused by the decreasing lower mass limit $m_{\rm min,Ceph}$ for Cepheids with $Z$. 
}
\label{fcephFrac_2p}
\end{figure*} \else \fi

\reff{fnCeph2} quantifies the mechanisms responsible for unbinding stars from clusters. 
For each cluster, we show two bars; the left bar represents ProCeps only, while the right bar represents 
lower mass stars only.
The fraction of lower mass stars which get unbound due to gas expulsion (blue bars) decreases monotonically with 
increasing cluster mass, which is a well known fact \citep[e.g.][]{Baumgardt2007}. 
However, ProCeps do not follow this trend, because they are preferentially retained in lower mass clusters 
due to dynamical mass segregation. 
Instead, the fraction of ProCeps which get unbound due to gas expulsion first increases with increasing $M_{\rm ecl}$, reaching 
a maximum of $0.52$ for $M_{\rm ecl} = 400 \Msun$ and further decreases as $M_{\rm ecl}$ increases. 
The maximum is caused by mass segregation of ProCeps in clusters with $M_{\rm ecl} \lesssim 400 \Msun$, 
which occurs before gas expulsion. 
The mechanism which unbinds most of the ProCeps is gas expulsion, 
with evaporation and ejections playing a secondary role; 
however none of these processes is negligible. 
Dynamical ejections unbind fewer ProCeps than evaporation. 
In total, $2$ to $10$ \% of ProCeps get ejected, where lower mass clusters produce a higher fraction of ejected stars. 
In contrast, ejections are negligible for lower mass stars.


\iffigscl
\begin{figure}
\includegraphics[width=\columnwidth]{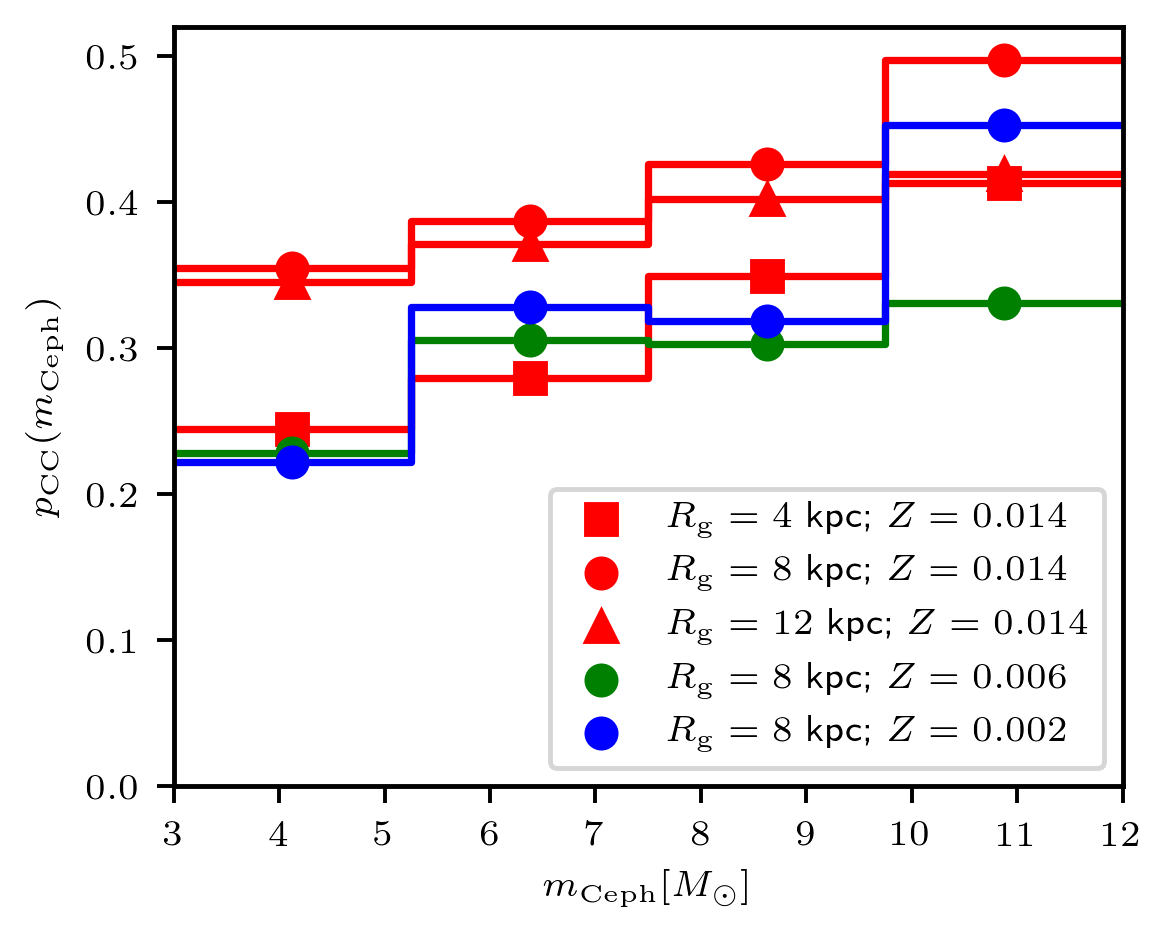}
\caption{
The fraction of cluster Cepheids $\pcciii$ as a function of the Cepheid mass $m_{\rm Ceph}$ for 
Cepheids originating from clusters at different Galactocentric radius $R_{\rm g}$ (indicated by symbols) and of different metallicity $Z$ (indicated by colour).
The plot represents Cepheids originating from the whole population of star clusters of the ECMF of \eq{eClusterIMF}.
Cepheids of higher mass have a slightly higher probability to be found in clusters.
}
\label{fmassCeph_frac}
\end{figure} \else \fi


\reff{fcluster_inOutRad} shows the fraction of Cepheids located in their birth clusters $\pccii$ 
as a function of the cluster age and mass. 
The values for clusters at $R_{\rm g} = 8 \Kpc$ are shown in the middle panel.
The histograms are produced by the Bayesian block method to take into account the non-uniformly sampled time when Cepheids occur. 
The main strengths of the method, which uses variable bin sizes, is its independence of the assumptions about the smoothness and shape of the 
function and the time resolution of the data \citep{Scargle2013,astropy2013}.
The plots feature all Cepheids formed within the simulations including UpCeps, 
which are present after time $t_{\rm He,end}$.
Clusters with $M_{\rm ecl} \gtrsim 800 \Msun$ have almost time independent $\pccii \approx 0.5$. 
Less massive clusters have lower $\pccii$, which decreases with time as these clusters dissolve more easily. 
The higher abundance $\eta$ of Cepheids in clusters of lower masses, which is due to mass segregation, is 
partially compensated due to faster dissolution of these clusters; $\pccii$ would be lower in the absence of mass segregation. 
The fraction $\pcci$ of cluster Cepheids averaged over time is shown in the left panel of \reff{fcephFrac_2p} (red circles). 

\iffigscl
\begin{figure}[h!]
\includegraphics[width=\columnwidth]{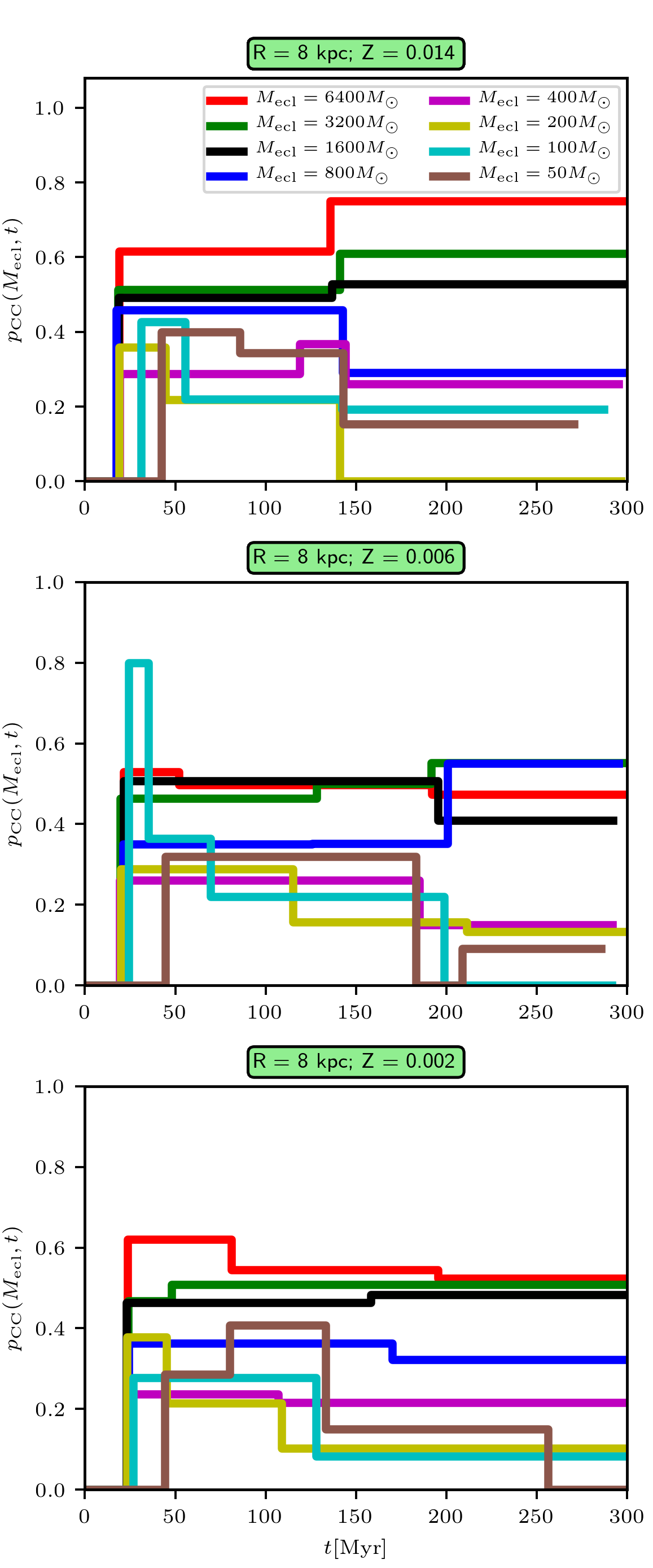}
\caption{
The fraction of cluster Cepheids $\pccii$ as a function of time for clusters of different metallicity:
$Z = 0.014$ (upper panel), $Z = 0.006$ (middle panel), $Z = 0.002$ (lower panel).
The cluster mass is indicated by colour. 
All clusters have $R_{\rm g} = 8 \Kpc$.
}
\label{fcluster_inOutZmet}
\end{figure} \else \fi

The earlier dissolution of lower mass clusters implies
that the Cepheids which occur in these clusters are on average 
of a larger mass $\overline{m}_{\rm Ceph}$ than those in more massive
clusters as shown in the right panel of \reff{fcephFrac_2p}, because lower
mass Cepheids occur at the time when the majority of lower
mass clusters have already dissolved, and the lower mass Cepheids have been released to the field.
Conversely, lower mass Cepheids are preferentially hosted by more massive clusters, which survive for longer. 
It appears that the dependence of $\overline{m}_{\rm Ceph}$ on $M_{\rm ecl}$ is robust 
because it is found in clusters at different orbital radii 
(red squares and triangles in \reff{fcephFrac_2p}) and metallicities (green and blue circles). 
Finally, the least massive clusters ($M_{\rm ecl} \lesssim 100 \Msun$)
have smaller $\overline{m}_{\rm Ceph}$ because their ability to host high-mass ProCeps is limited by the IMF
(cf. $m_{\rm max}$ in \reft{tsimList}). 

The occurrence of Cepheids in clusters therefore depends on several aspects. 
In particular, we note the dichotomy of 
$\pccii$ with age for high and low cluster masses, where $M_{\rm ecl} \approx 800 \Msun$ separates the two regions. 


The properties of ProCeps and Cepheids for the lower mass clusters show significant differences between individual realisations 
for clusters of the same mass. 
As an example, clusters of mass $200 \Msun$ contain on average only $3.5$ ProCeps (c.f. $N_{\rm PC}$ and $\rm{Nmod}$ in \reft{tsimList}), 
which implies small number statistics for Cepheids in low mass clusters. 
Some low mass clusters do not follow the evolutionary path described in this section, where interaction of each individual ProCep 
with another ProCep or lower mass star decides whether these stars are retained in the cluster or not. 
For example, all ProCeps can be ejected from the cluster in a single dynamic encounter occurring before gas expulsion (i.e. $\lesssim 0.6 \Myr$). 
The coarseness is absent in more massive clusters ($6400 \Msun$ clusters have $\approx 125$ ProCeps). 
This should be taken into account when comparing present theoretical findings with observations. 
Our results describe averages over a large cluster sample, 
so they are representative, for example, for a cluster population of a galaxy rather than for the individual cluster. 

So far, we aimed at theoretically understanding of the fraction of Cepheids in clusters with little emphasis on quantities which can be directly 
compared with observations. 
One quantity of more observational interest is the fraction of Cepheids of given mass, $m_{\rm Ceph}$, which is located in clusters (\reff{fmassCeph_frac}). 
In this plot, the Cepheids at any mass bin originate from the whole star cluster population of the ECMF of \eq{eClusterIMF}. 
The fraction $\pcciii$ increases with $m_{\rm Ceph}$ because more massive Cepheids occur earlier, when more lower mass star clusters are still gravitationally bound 
(it is the same effect which is responsible for the higher mean Cepheid mass $\overline{m}_{\rm Ceph}$ in lower mass clusters as seen in the 
right panel of \reff{fcephFrac_2p}).


\subsection{The role of galactocentric radius}

\label{ssGalRad}

The clusters at $R_{\rm g} = 4 \Kpc$ (upper panel of \reff{fcluster_inOutRad}) and $R_{\rm g} = 12 \Kpc$ 
(lower panel) follow the same trends in $\pccii$
as the cluster at $R_{\rm g} = 8 \Kpc$ (middle panel), which was described in the previous section:
The value of $\pccii$ decreases with time for lower mass clusters because they dissolve faster than  
more massive clusters (c.f. \reff{fdissolved}), which have $\pccii$ almost independent of time. 
We note that Cepheids occur non-uniformly with time, with $\approx 80$\% of all Cepheids occurring before $t_{\rm He,end} \approx 150 \Myr$ (for $Z = 0.014$). 
This influences their statistics, which is relatively good at $\lesssim 150 \Myr$, but relatively poor afterwards, when only a small number of  
Cepheids can cause visible difference. 
While most of the clusters follow the expected trend of having larger $p_{\rm CC} (M_{\rm ecl}, t)$ at a larger galactocentric radius, 
there are several exceptions because of low number statistics. 
An example are the $M_{\rm ecl} = 100 \Msun$ clusters (cyan lines), which have lower $p_{\rm CC} (M_{\rm ecl}, t \gtrsim 150 \Myr)$ for the $R_{\rm g} = 12 \Kpc$ 
models than for the $R_{\rm g} = 8 \Kpc$ models.

The time-averaged fraction of Cepheids in clusters for $R_{\rm g} = 4 \Kpc$ (red squares), $R_{\rm g} = 8 \Kpc$ (red circles)
and $R_{\rm g} = 12 \Kpc$ (red triangles) is shown in the left panel of \reff{fcephFrac_2p}. 
For lower mass clusters ($M_{\rm ecl} \lesssim 400 \Msun$), $\pcci$ generally increases with increasing $R_{\rm g}$ as the galactic tidal field becomes less important.
Averaged over all the clusters with $M_{\rm ecl} \lesssim 400 \Msun$, 
$f_{\rm CC} = 0.17$ for $R_{\rm g} = 4 \Kpc$, $f_{\rm CC} = 0.26$ for $R_{\rm g} = 8 \Kpc$, 
and $f_{\rm CC} = 0.27$ for $R_{\rm g} = 12 \Kpc$. 
On the other hand, more massive clusters ($M_{\rm ecl} \gtrsim 800 \Msun$), 
have $f_{\rm CC} \approx 0.5$ almost independently of $R_{\rm g}$ because they are not much affected by 
the tidal field of the galaxy.

\subsection{The role of metallicity}

\label{ssZmet}


\reff{fcluster_inOutZmet} shows the time evolution of $\pccii$ for clusters of different metallicity. 
At a lower metallicity, Cepheids occur at a lower mass $m_{\rm min,Ceph}$ and live longer (their $t_{\rm He,end}$ is larger). 
Consequently, $\pccii$ reaches lower values near $t_{\rm He,end}$ for clusters with lower metallicity because 
the clusters have more time to evaporate or dissolve; 
this trend is more pronounced in lower mass clusters ($p_{\rm CC} (M_{\rm ecl}, t_{\rm He,end}) \lesssim 0.1$ for $Z = 0.002$) 
because they release stars to the field faster. 
When integrated over time (left panel of \reff{fcephFrac_2p}), 
the value of $f_{\rm CC}$ averaged over clusters with $M_{\rm ecl} \lesssim 400 \Msun$ decreases from $f_{\rm CC} = 0.26$ 
for $Z = 0.014$ via $f_{\rm CC} = 0.24$ for $Z = 0.006$ to $f_{\rm CC} = 0.18$ for $Z = 0.002$. 
Clusters with $M_{\rm ecl} \gtrsim 800 \Msun$ have $f_{\rm CC} \approx 0.5$ with no dependence on $Z$.
This shows that the highest sensitivity of $\pcci$ 
on metallicity is for lower mass clusters ($M_{\rm ecl} \lesssim 800 \Msun$).

This result is the consequence of the longer life-span of Cepheids at a lower metallicity, at which 
time more lower mass clusters is dissolved, instead of differences in internal dynamics for clusters of 
a lower metallicity. 
This can be deduced from \reff{fdissolved}, 
where the fraction of gravitationally bound clusters at a given time, $f_{\rm cl,bound}^{\rm sim}$, is 
almost independent of metallicity (c.f. the solid, dashed and dotted lines). 
This result is also supported by \citet{Hurley2004}, who find a rather minor influence of metallicity on star cluster evolution even 
for a metallicity substantially lower than considered here ($Z = 10^{-4}$).
Additionally, the lower mass of Cepheids at a lower metallicity (right panel of \reff{fcephFrac_2p}) 
means that their progenitors need a longer time-scale for mass segregation, and are more likely 
to escape the cluster during gas expulsion, further reducing the population of cluster Cepheids. 

\iffigscl
\begin{figure*}
\includegraphics[width=\textwidth]{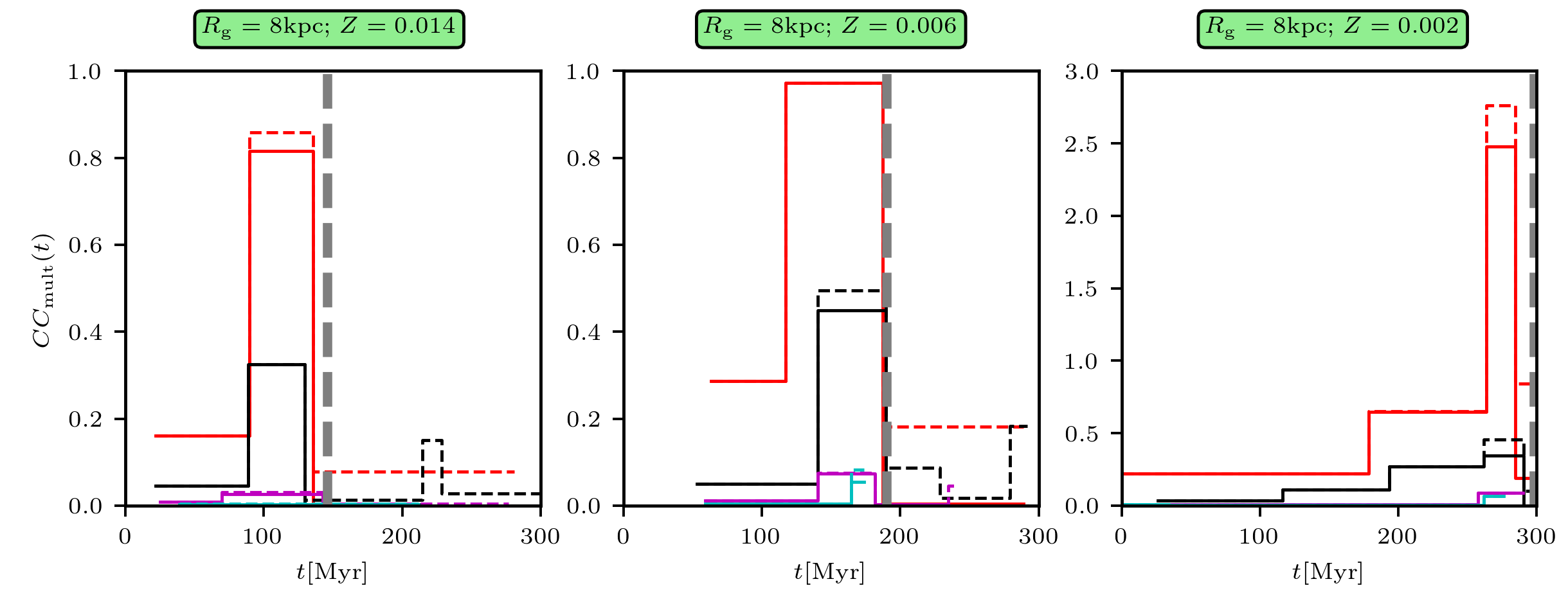}
\caption{
The time dependence of $CC_{\rm mult}$ for CanCeps (solid lines) and all Cepheids (dashed lines) in clusters of 
different mass as shown by the colour.
The grey vertical dashed lines indicate the time $t_{\rm He,end}$ for the given metallicity. 
The histograms are binned by the Bayesian block method. 
The lines are shown for the interval marked by the occurrence of the first and last Cepheid in clusters in present models 
(the discretisation to individual Cepheids is the reason why all models do not reach exactly $300 \Myr$).
}
\label{foccur}
\end{figure*} \else \fi

\iffigscl
\begin{figure*}
\includegraphics[width=\textwidth]{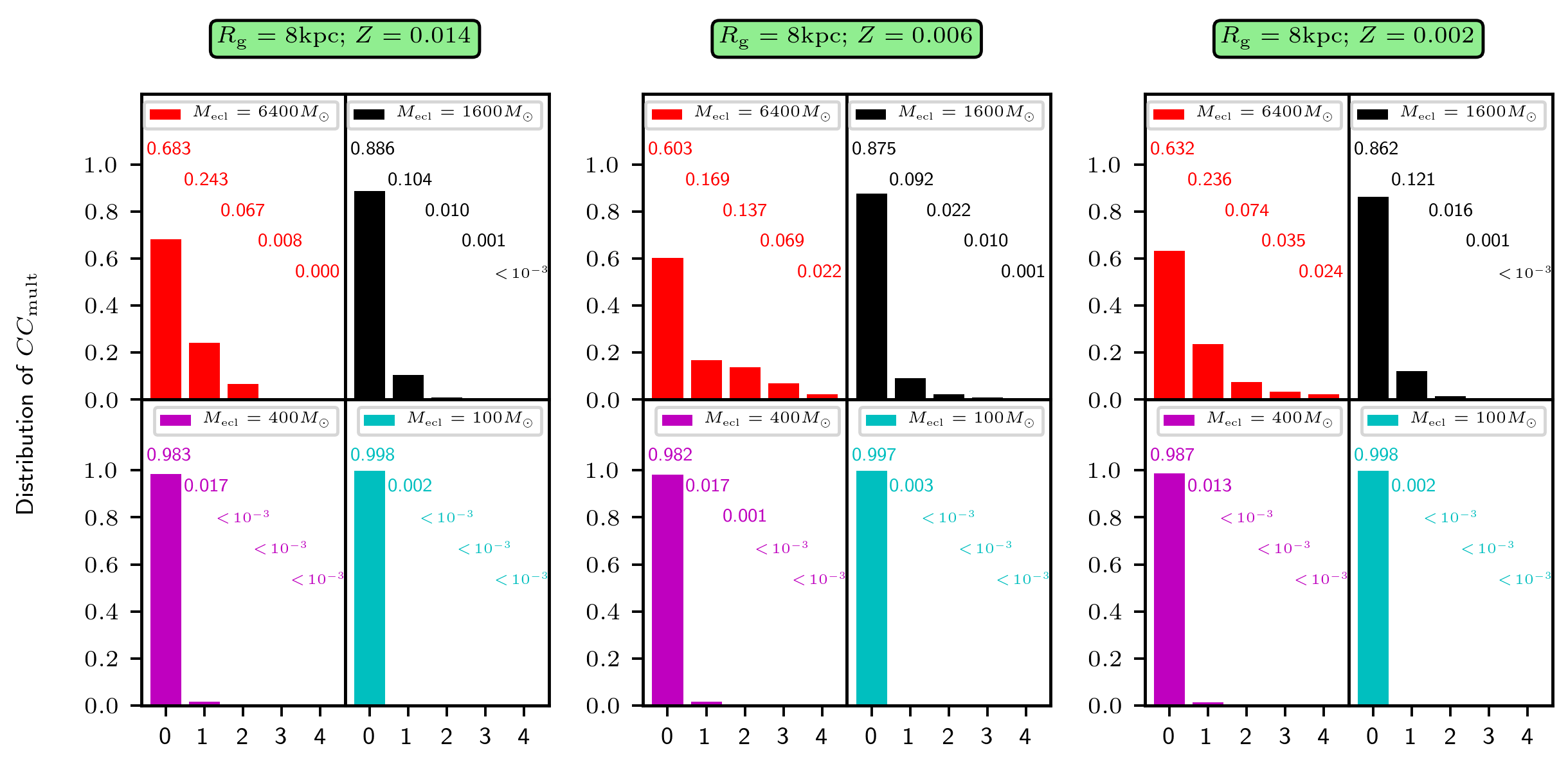}
\caption{
Normalised histograms of the number of Cepheids located in the same cluster, $CC_{\rm mult}$, at a given time as a function of cluster mass and metallicity. 
The numbers above the columns indicate the number fraction of clusters which contained zero to four Cepheids at the same time. 
The Cepheids taken into account are the CanCeps only. 
None of the clusters contained more than six Cepheids at the same time.
}
\label{fCCMult}
\end{figure*} \else \fi

\iffigscl
\begin{figure}
\includegraphics[width=\columnwidth]{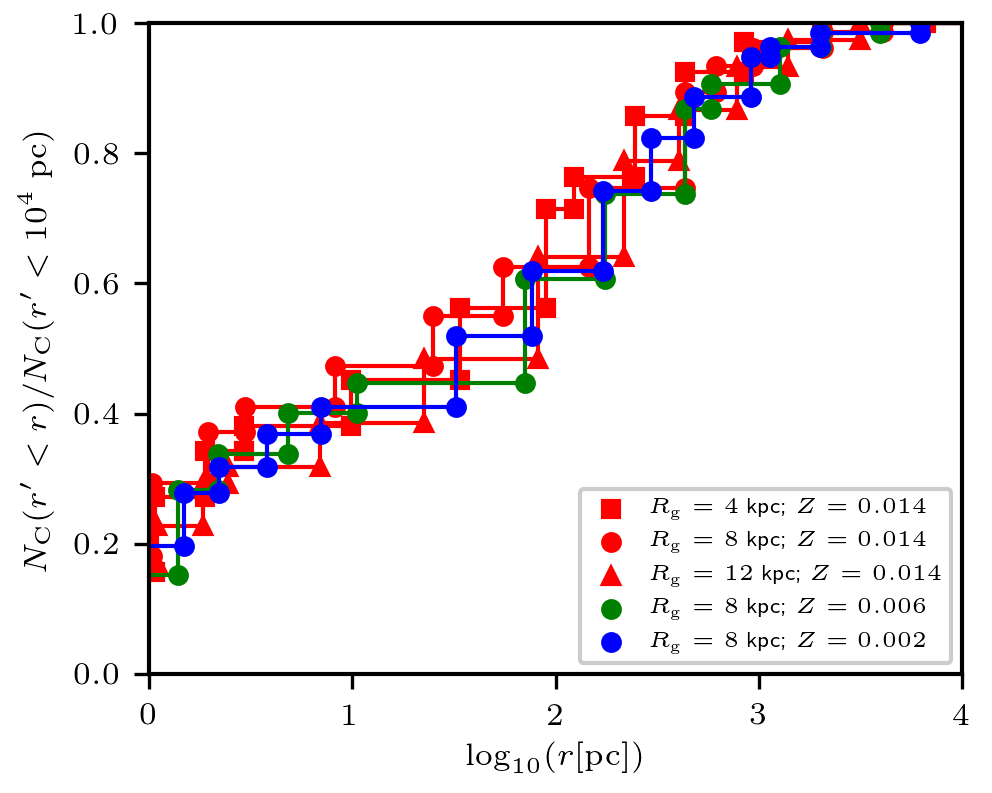}
\caption{
The cumulative distribution of Cepheids according to the distance $r$ to the density centre of their birth cluster for the whole population 
of star clusters from the ECMF of \eq{eClusterIMF}.
We plot only the Cepheids whose birth cluster is still gravitationally bound at the time of the occurrence of the Cepheid.
}
\label{fdist_icmf}
\end{figure} \else \fi

\subsection{The number of Cepheids hosted by a cluster at the same time}

\label{ssMultCep}

To interpret the results, we remember that $M_{\rm ecl}$ is the birth mass of the cluster. 
The present mass of the cluster hosting Cepheids is typically by a factor of two to three lower than this, depending on its age and initial $M_{\rm ecl}$.

\reff{foccur} shows the time-averaged number of Cepheids, $CC_{\rm mult}$, occurring at the same time within the same cluster 
as a function of the cluster age for clusters of four different masses.
Since the number of Cepheids in the same cluster rapidly varies with time around small integer numbers, 
forming a teeth-like curve, 
we resort to taking a time-average of this quantity.
The histogram of the number of Cepheids is discussed below (see also \reff{fCCMult} below). 
The value of $CC_{\rm mult}$ increases with time up to $t_{\rm He,end}$ (\reff{foccur}; the grey dashed line) because lower mass Cepheids, 
which occur later are more abundant
(according to the adopted model for the IMF), and they spend a longer time 
within the instability strip; this occurs despite the continuing cluster dissolution with progressing time. 
More massive clusters, which contain more ProCeps, have higher $CC_{\rm mult}$. 
$CC_{\rm mult}$ also increases with decreasing metallicity (particularly before $t_{\rm He,end}$) because of the lower minimum mass of Cepheids 
(and thus their higher abundance) in this environment (right panel of \reff{foccur}).
After the expected life-time of ProCeps $t_{\rm He,end}$, $CC_{\rm mult}$ suddenly drops, but it is non-zero; these Cepheids originate from UpCeps, 
which are shown by dashed lines (a more detailed study of UpCeps is available in Paper II).

How many Cepheids do we expect to see in a particular cluster at the same time? 
To answer this question, we count the number of Cepheids in each of our simulated clusters every $30 \Kyr$ during the time interval 
of occurrence of CanCeps, i.e. between $15 \Myr$ and $t_{\rm He,end}$. 
An histogram of this count is shown in \reff{fCCMult}.
We note first that the histogram pertains to observations of a single star cluster, not the whole population of star clusters. 
For example, the probability of finding at least one Cepheid in a cluster with a birth stellar mass of $6400 \Msun$ is $p = 0.317$, while 
for an $400 \Msun$ cluster the probability is only $p = 0.017$. 
However, the galaxy forms $\approx 16 \times$ more $400 \Msun$ clusters than $6400 \Msun$ clusters, so the probability of finding at least one Cepheid 
among $400 \Msun$ clusters is not that different as among $6400 \Msun$ clusters (e.g. $p = 0.017 \times 16 = 0.272$ vs. $p = 0.317$). 

The probability of finding more Cepheids at the same time decreases with the number of Cepheids we intend to find, and also with decreasing cluster mass (\reff{fCCMult}). 
Clusters harbouring two and more Cepheids at the same time are usually the more massive ones, with mass larger than $400 \Msun$. 
Clusters with initial mass $\lesssim 400 \Msun$ almost never host more than one Cepheid. 
The highest value of $CC_{\rm mult}$ found in our simulations is five, and this occurred for the most massive ($M_{\rm ecl} = 6400 \Msun$), 
and the least metallic ($Z = 0.002$) cluster. 
The event of $CC_{\rm mult} = 5$ occurred only once in more than $2500$ simulations in total, so observations of clusters of mass below $\approx 6400 \Msun$ 
with $CC_{\rm mult} > 5$ are likely to be rare. This low probability for two or more Cepheids to occur simultaneously in a given cluster 
agrees very well with the observed statistics based on the MW, M31, and the majority of Cepheid hosting clusters in the LMC and SMC \citep{Anderson2018}. 

However, some LMC clusters hosting many Cepheids are remarkable exceptions to this general rule, 
notably NGC~1866 and NGC~2031 that each host $\approx 20$ or more Cepheids \citep{Welch1993,Testa2007}. 
We assess whether the occurrence of such clusters is consistent with our simulations in the following way. We assume that a Cepheid in a cluster occurs at a time which is independent of the occurrence of other Cepheids, and 
that the probability $p_{\rm ioc}$ to find a given ProCep as a Cepheid at any time is substantially smaller than 1. 
These assumptions imply Poisson distribution for Cepheids with mean value 
\begin{equation}
\lambda = p_{\rm ioc} N_{\rm Ceph},
\label{ePoisson}
\end{equation}
where $N_{\rm Ceph}$ is the number of Cepheids which occurred during the cluster's life-time. 
From the values in the middle panel of \reff{fCCMult} (metallicity similar to NGC~1866), 
we obtain $\lambda = 0.97$ and $\lambda = 0.22$ for models with $M_{\rm ecl} = 6400 \Msun$ 
and $M_{\rm ecl} = 1600 \Msun$, respectively. 
These values are consistent with the definition of $\lambda$, which is linearly proportional to $N_{\rm Ceph}$. When scaling $\lambda$ from the $M_{\rm ecl} = 1600 \Msun$ clusters to the $M_{\rm ecl} = 6400 \Msun$ clusters, we would 
obtain $\lambda = 0.88$ for $M_{\rm ecl} = 6400 \Msun$ clusters. 
Averaging these two points yields $\lambda \approx 0.93$ for $M_{\rm ecl} = 6400 \Msun$ clusters. Adopting a mass of $1.35 \times 10^5 \Msun$ for NGC\,1866 \citep{Fischer1992}, \eq{ePoisson} yields $\lambda = 19.6$. 
Thus, the Poisson distribution yields an expected 20 Cepheids, which is not too far from the known 24 Cepheids in NGC\,1866 \citep{Welch1993} whose 
simultaneous occurrence has a probability of $\approx 13\,\%$.

Thus, the wealth of Cepheid members of these two LMC clusters is explained by their high mass, low metallicity (more numerous lower mass stars can become Cepheids), and a cluster age consistent with the occurrence of comparatively old short-period (low-mass) Cepheids. 
For reference, the mean pulsation periods of Cepheids in NGC\,1866  and NGC\,2103 reported by \citet{Testa2007} are $2.9\,$d and $3.2$\,d, 
respectively, corresponding to ages of $\approx 100$\,Myr assuming period-age relations without rotation \citep{Bono2005} and approximately $200$\,Myr for 
period-age relations assuming average main sequence rotation rates \citep{Anderson2016}. 



\subsection{Cepheids in the outskirts of star clusters}

\label{ssOutskirts}


\reff{fdist_icmf} shows the cumulative distribution of Cepheids according to the distance $r$ from 
the density centre of their birth cluster. 
Each star is plotted at the time when it becomes a Cepheid, which means that the stars are shown at different ages. 
This enables a direct comparison with observations.
Approximately 20 \% of Cepheids whose birth cluster is still gravitationally bound are located at the distance from $10 \Pc$ to $50 \Pc$. 
Interestingly, \citet{Anderson2013} find that 5 of 23 ($22$ \%) cluster Cepheids in their sample are located at this distance from clusters. 
Recent searches for Cepheids in MW clusters have primarily reported such new possible ``halo'' members \citep{Medina2021,Zhou2021}. 
We find approximately $5\%$ of Cepheids located far away ($\gtrsim 1 \Kpc$) from their birth clusters as a result of dynamical ejection.
The distribution of Cepheids according to $r$ is independent of the galactocentric radius $R_{\rm g}$ or metallicity $Z$.

\section{The fraction of Cepheids in star clusters}

\label{sFracCeph}

In this Section, we provide the likely upper and lower estimate on $f_{\rm CC}$.
The upper estimate on $f_{\rm CC}$ is taken from present simulations, where the main mechanisms responsible for cluster dissolution are the early gas expulsion and 
the tidal field of the galaxy. 
However, star clusters are subjected to additional dissolution mechanisms, e.g. encounters with giant molecular 
clouds (GMCs; \citealt{Spitzer1958,Terlevich1987,Theuns1992,Jerabkova2021}), which is not included in present simulations. 
The lower estimate on $f_{\rm CC}$ is based on the difference between the population of observed open star clusters in the Galaxy 
and the population of bound star clusters in our models. 
Note that we do not a priori assume the physical mechanism for additional dissolution, we only assume the survival rate of star clusters.

We consider four different models of the galactic cluster population. 
In the first three models, 
we assume that at any galactocentric radius $R_{\rm g}$, Cepheids are formed in star clusters of the ECMF and initial mass range as given in \refs{ssECMF}. 
Since the cluster number density often varies non-trivially as a function 
of the galactocentric radius with a significant fraction of star formation sometimes 
located in ring-like structures
\citep[e.g.][]{Lewis2015},
we assume rather extreme radial distributions of star clusters over the galaxy to constrain the possible variations in $f_{\rm CC}$: 
(i) all star clusters are located at 
the same galactocentric radius of $R_{\rm g} = 4 \Kpc$; (ii) all star clusters are located at 
the same galactocentric radius of $R_{\rm g} = 12 \Kpc$; (iii) star clusters 
follow the usual exponential stellar surface density profile of spiral galaxies in the form of $\Sigma_{\rm ecl} \propto \exp{(-R_{\rm g}/R_{\rm d})}$ with 
the disc scale length $R_{\rm d} = 2.5 \Kpc$ \citep{Binney2008}. 

The fourth model (iv) is based on the idea of \citet{PflammAltenburg2008}, where the maximum cluster mass $M_{\rm ecl,max}$
decreases with increasing $R_{\rm g}$ in disc galaxies as $M_{\rm ecl,max}(R_{\rm g}) = M_{\rm ecl,max,0} \exp{(-\gamma R_{\rm g}/R_{\rm d})}$ 
because it is set by the local gas surface density. 
This dependence of the cluster masses on galactocentric radius has been documented for the late-type galaxy M33 by 
\citet{PflammAltenburg2013}, which also demonstrates the distribution of young clusters to be non-stochastic.
Following \citet{PflammAltenburg2008}, we adopt the slope $\gamma = 1.5$, the disc scale length $R_{\rm d} = 4 \Kpc$, 
and the maximum cluster mass which the galaxy can form $M_{\rm ecl,max,0} = 8.10^4 \Msun$ \citep{Weidner2004}. 
The final value of $f_{\rm CC}$ is calculated by weighting the ECMF of \eq{eClusterIMF} (in the interval of $50 \Msun$ to $M_{\rm ecl,max}(R_{\rm g})$) 
by the number of Cepheids forming in the particular annulus, which is proportional to $2 \pi R_{\rm g} \exp{(-\gamma R_{\rm g}/R_{\rm d})}$.

\begin{table*}
\begin{tabular}{c|cccc|cccc}
$Z$ & \multicolumn{4}{c}{lower $f_{\rm CC}$} & \multicolumn{4}{c}{upper $f_{\rm CC}$} \\
\hline
 & (i) & (ii) & (iii) & (iv) &   (i) & (ii) & (iii) & (iv) \\
\hline
0.014 & 0.026 & 0.015 & 0.024 & 0.021 & 0.30 & 0.36 & 0.33 & 0.34 \\
0.006 & 0.024 & 0.014 & 0.023 & 0.019 & 0.27 & 0.32 & 0.29 & 0.30 \\
0.002 & 0.024 & 0.014 & 0.022 & 0.017 & 0.25 & 0.29 & 0.27 & 0.27
\end{tabular}
\caption{The total fraction $f_{\rm CC}$ of Cepheids in star clusters for the whole galaxy derived 
under different assumptions about the cluster distribution within the galaxy (indicated by Roman numerals; they are defined in \refs{sFracCeph}), 
cluster dissolution processes and metallicity $Z$. 
Columns 2-5 assume pronounced cluster dissolution with the fraction 
$f_{\rm cl,bound}^{\rm obs}(t)$ of 
clusters remaining gravitationally bound by the age $t$ given by \eq{eDissolution}. 
This is likely the lower estimate on $f_{\rm CC}$. 
Columns 6-9 assume cluster dissolution by the galactic tidal field only, which provides an upper estimate on $f_{\rm CC}$. 
}
\label{tResults}
\end{table*}

\subsection{The upper estimate on $f_{\rm CC}$}

\label{ssFracCephUp}

For the cluster distributions (i), (ii), (iii) and (iv) and for solar metallicity, we obtain $f_{\rm CC} = 0.30$, $f_{\rm CC} = 0.36$, $f_{\rm CC} = 0.33$, 
and $f_{\rm CC} = 0.34$, respectively (c.f. the sixth to ninth column of \reft{tResults}).  
For the sub-solar metallicities, $Z = 0.006$ and $Z = 0.002$, we have models only for $R_{\rm g} = 8 \Kpc$, for which we obtain 
$f_{\rm CC} = 0.35$ and $f_{\rm CC} = 0.32$, respectively. 
Since metallicity has a minor influence on dynamical evolution of our models (c.f. \refs{ssZmet}), we estimate $f_{\rm CC}$ 
in sub-solar metallicity clusters by scaling their values from $R_{\rm g} = 8 \Kpc$ to $R_{\rm g} = 4 \Kpc$ and $R_{\rm g} = 12 \Kpc$ by the same 
factors as is the scaling for the model with $Z = 0.014$. 
We obtain $f_{\rm CC}$ in the range from $0.27$ to $0.32$ for $Z = 0.006$, and $f_{\rm CC}$ in the range from $0.25$ to $0.29$ for $Z = 0.002$ (\reft{tResults}). 

\subsection{The lower estimate on $f_{\rm CC}$}

\label{ssFracCephDown}

In the solar vicinity, only $\approx 10$\% of embedded star clusters survive to the age of $10 \Myr$, and $\approx 4$\% 
of embedded star clusters survive to the age of $100 \Myr$ \citep{Lada2003}, which is attributed mainly to 
encounters with GMCs. 
The observed fraction of surviving clusters is smaller than what we obtain in our simulations 
(c.f. \reff{fdissolved}), which implies that more Cepheids are released to the field in the real galaxy.
However, the inclusion of more realistic (i.e. non-spherically symmetric and evolving with time) GMCs 
to this type of simulation would be a non-trivial task, which to our knowledge has not been accomplished yet, 
so we resort to a coarse approximation to obtain an order of magnitude estimate. 

Based on \citet{Lada2003}, we approximate the fraction of star clusters which remain gravitationally bound at age $t$ by 
\begin{equation}
f_{\rm cl,bound}^{\rm obs}(t) = 0.1 (t/10 \Myr)^{-2/5}.
\label{eDissolution}
\end{equation}
We assume that \eq{eDissolution} is independent of the cluster mass, and of $R_{\rm g}$.
Then, for each Cepheid in a cluster (which occurs at the cluster age $t$), we measure  
the fraction of bound clusters $f_{\rm cl,bound}^{\rm sim}(t)$ in our simulations. 
In the real galaxy, there is $f_{\rm cl,bound}^{\rm obs}(t)/f_{\rm cl,bound}^{\rm sim}(t)$ times Cepheids in clusters at $t$, so we take   
$\min(f_{\rm cl,bound}^{\rm obs}(t)/f_{\rm cl,bound}^{\rm sim}(t), 1)$ of each Cepheid as located within a star cluster.

The fraction of cluster Cepheids obtained by this correction for the additional dissolution processes is shown in the second to fifth column 
of \reft{tResults}. 
It is by a factor of $10$ smaller than without the correction (sixth to ninth column), 
which indicates that molecular clouds or a similar disruptive mechanism play likely a crucial role in setting the value of $f_{\rm CC}$. 
These values of $f_{\rm CC}$ are comparable to what is observed in the galaxy M~31 ($f_{\rm CC} = 0.025$; \citealt{Anderson2018}).

\section{Discussion}

\label{sDiscussion}

\subsection{Gas expulsion and mass segregation}

\label{ssGasExpulsion}

The most important parameters of gas expulsion which influence cluster dynamics is the star formation efficiency $\sfe$ and the gas expulsion time-scale $\tau_{\rm M}$ 
\citep{Baumgardt2007}. 
Although we adopt what appear to be their most likely values \citep{Kroupa2001b,Banerjee2013,Megeath2016}, we discuss the possible influence on $f_{\rm CC}$ in case that 
these parameters are markedly different. 
For SFE higher than $1/2$, gas expulsion unbinds only a small fraction of stars from the cluster. 
A small fraction of stars is released also when $\tau_{\rm M} \gg t_{\rm cr}$. 
Nevertheless, lower mass clusters dissolve by $300 \Myr$ due to cluster dynamics only \citep{FuenteMarcos2004}, 
while the more massive clusters are likely to retain the majority of their ProCeps.
In these cases, we expect $f_{\rm CC}$ to be somewhat larger than in our models. 

On the other hand, if $\sfe \lesssim 1/3$ and $\tau_{\rm M} \lesssim t_{\rm cr}$, the majority of clusters is quickly dispersed forming unbound OB associations
with a possible handful of small bound clusters, which are remainders from the most massive and compact clusters \citep{Tutukov1978}. 
In this case, $f_{\rm CC}$ would be lower than in our models, possibly similar to our lower estimate on the $f_{\rm CC}$.

The present models started as non-mass segregated, but the lower mass clusters ($M_{\rm ecl} \lesssim 200 \Msun$) developed mass segregation dynamically 
before the onset of gas expulsion at $t_{\rm d} = 0.6 \Myr$. 
However, the value of $t_{\rm d}$ is not well constrained. 
If gas expulsion started earlier (later), mass segregation will develop in clusters less (more) massive than $200 \Msun$.

Mass segregation (either primordial or induced by dynamics) impacts $\pcci$ in several ways. 
As shown in \refs{ssCephDyn}, mass segregation retains more ProCeps in the cluster in comparison to lower mass stars, which tends to increase $f_{\rm CC}$ in total. 
On the other hand, this effect is to some degree counteracted by stellar evolution when mass segregated massive stars explode as supernovae 
near the cluster centre, causing more pronounced cluster expansion, unbinds more stars and results in faster cluster dissolution \citep{Brinkmann2017}, 
which would decrease $f_{\rm CC}$.
The final outcome of the interplay of these two processes, which likely depends on both the cluster mass and Cepheid mass, 
is difficult to predict without further N-body calculations.


\subsection{Initial binary fraction\label{ssInitialCond}}

The assumption of $100\%$ binary fraction under the exemption of initial triple systems may have significant impact on our results. 
Paper\,II in this series studies in detail the dynamics of multiple systems in our simulation. 
It suffices here to state that our simulations suggest that a significant fraction (possibly the majority) 
of mid-B stars form in triples and quadruple systems instead of binaries. 
The presence of primordial triples can cause faster mass segregation (because these subsystems are more massive than single stars or binaries) 
and preserve more ProCeps in clusters increasing $f_{\rm CC}$; 
on the other hand, the dynamical interactions within subsystems and between subsystems and cluster members are likely to cause more ejections, decreasing $f_{\rm CC}$. 
Additional simulations spanning a wide parameter range of initial conditions would be required to study these effects in further detail.

\subsection{The external potential of the galaxy}

\label{ssGalPotential}

The gravitational potential of the galaxy was modelled using the approximation of \citet{Allen1991}, even though more recent approximations are available 
\citep[e.g.][]{Irrgang2013,Pouliasis2017}. 
We argue that the details of the potential are not critical for our results because the main influence of the potential is to define the tidal radius, 
beyond which the stars are removed from the cluster, thus setting the rate of evaporation. 
The clusters are on circular orbits, so the tidal radius does not change as the cluster revolves the galaxy (apart from cluster mass loss due to evaporation). 
A cluster of mass $M_{\rm cl}$ on a circular orbit of radius $R_{\rm g}$ has a tidal radius given by
\begin{equation}
r_{\rm t} = R_{\rm g} \left(\frac{M_{\rm cl}}{4 M_{\rm glx}} \frac{\gamma^2}{(\gamma^2 - 1)} \right)^{1/3},
\end{equation}
\citep[][their eqs. 3.80 and 8.106]{Binney2008}, where 
$M_{\rm glx}$ is the mass of the galaxy within the cluster orbit and $\gamma = 2 \Omega/\kappa$ is the ratio of the orbital and epicyclic frequency. 
Since $M_{\rm glx}$ follows from the magnitude of the rotational curve and $\gamma$ follows from its slope, which 
is flat for most of spiral galaxies \citep{Bosma1981,Rubin1982} implying $\gamma \approx 1.4$, the value of the tidal radius is insensitive 
of given galaxy model for clusters located in the stellar disc. 
We note that in another theories of gravity (e.g. MOND), cluster evaporation can occur on a different time-scale (faster in the case of MOND) and the tidal structure 
around the cluster can be asymmetric (Kroupa et al. in preparation).

This study aims at general spiral galaxies, which show a large variability in their gravitational potentials, so using a model 
particularly tuned for the Galaxy would not provide a general result.

\subsection{The role of molecular clouds\label{ssMoleCloud}}

The lower estimate on $f_{\rm CC}$ in our models (\reft{tResults}) is similar to the observed fraction of cluster Cepheids in
M~31, $f_{\rm CC,M31} = 0.025$, \citep{Senchyna2015,Anderson2018}.  However, significant variations in $f_{\rm CC}$ can exist among galaxies,
and $f_{\rm CC} \approx 0.07 - 0.11$ was estimated in the LMC, SMC, and the Milky Way \citep[cf. their Sect. 3.1]{Anderson2018}.

We here provide a tentative update on the Milky Way value of $f_{\rm CC}$ using the high-quality astrometry from
the Gaia mission's Early Data Release 3 \citep[]{Gaiac2021}, updated literature on Galactic open clusters
\citep{Loktin2017,Cantat2018,Cantat2020,Ferreira2021,Hunt2021,Dias2021}, recent investigations of Cepheid membership in Galactic clusters
\citep{Breuval2020,Zhou2021,Medina2021}, and the most extensive literature on Galactic Cepheids from the General Catalog of Variable
Stars \citep{GCVS2017}, the Variable Star Index \citep{VSX}, and the re-classification of MW Cepheids by \citet{Ripepi2019}. A more detailed
investigation of this fraction will be presented separately (Cruz Reyes et al. in prep.). We limit our estimate to a volume of $2$ kpc to
strike a useful balance between statistics and completeness of Cepheids and clusters. Following several quality cuts and detailed inspection
of light curves and other available data, we find $217$ classical Cepheids within $2$ kpc. Of these, $15$ are bona fide members of star
clusters. We thus find $f_{\rm CC,MW}=15/217=0.069$, somewhat lower than the previous value of $11/130=0.085$ in \citet{Anderson2018}. On the
one hand, the decrease in $f_{\rm CC,MW}$ is driven by the greater number of Cepheids that can be placed within $2$ kpc thanks to \textit{Gaia}
astrometry. On the other hand, the increase in known clusters resulted in several new cases as well as a reassessment of cluster membership
of some other cases. However, it is likely that the list of known clusters within $2$ kpc remains significantly less complete than the list of
known Cepheids, as evidenced by the recent discovery of more than 600 clusters, mostly at distances beyond 1 kpc \citep{Castro2021}. We therefore
consider our updated estimate of $f_{\rm CC,MW}=15/217=0.069$ a lower limit to the true fraction of cluster Cepheids within 2 kpc from the Sun.

Such large differences between different galaxies cannot be fully explained by lower metallicity ($Z=0.006$ and $Z=0.002$ was used for the LMC and SMC, respectively)
nor by a different strength of the galactic tidal field. 
As shown in \refs{ssGalRad} and \refs{ssZmet}, decreasing metallicity from $Z = 0.014$ to $Z = 0.002$ decreases $f_{\rm CC}$ only by a factor of $1.2$, 
and increasing galactocentric radius from $R_{\rm g} = 4 \Kpc$ to $R_{\rm g} = 12 \Kpc$ increases $f_{\rm CC}$ also by a factor of $1.2$
\footnote{The present clusters are calculated for the external gravitational field of
the Galaxy.  If the clusters were positioned in the LMC and SMC, their tidal radii would be only marginally larger by a factor of $1.3$ and $1.1$,
respectively, (the estimate is based on rotation curves from \citealt{Stanimirovic2004} and \citealt{Marel2014}), than if they were located in
the Galaxy at $R_{\rm g} = 8 \Kpc$.  Given the modest variation of $f_{\rm CC}$ on $R_{\rm g}$, we do not expect that the tidal field of the
Magellanic Clouds causes the large difference in the observed value of $f_{\rm CC}$.}.  
A possible explanation of the difference of $f_{\rm
CC}$ for different galaxies might stem from different conditions in the interstellar medium in these galaxies, which forms molecular clouds
of different masses and/or on different time-scales, thus having a different impact on star clusters via the gravitational force.  If true,
this would imply a bias depending on the gaseous content of the galaxy.



\section{Conclusions}

\label{sSummary}




We investigated the dynamical evolution of Cepheids in open star clusters with the main aim to estimate the fraction of Cepheids located in clusters and the field. 
We focused on the influence of the initial cluster mass, orbital radius within its galaxy and metallicity. 
The stellar masses follow a realistic initial mass function and all stars are assumed to form as binaries.
At a young age, the clusters expel their non-star forming gas; this unbinds a significant fraction of stars from the clusters.
All the results are obtained directly from simulations.

Lower mass clusters dissolve more easily in the tidal field of the galaxy, releasing a higher fraction of their stars to the field than more massive clusters. 
For more massive stars, including ProCeps, this process is partially compensated by the rapid mass segregation in very low mass clusters ($M_{\rm ecl} \lesssim 200 \Msun$) 
occurring before gas expulsion. 
The fraction of Cepheids located in clusters increases with the cluster mass from $\approx 0.25$ (for $M_{\rm ecl} = 50 \Msun$) to 
$\approx 0.60$ (for $M_{\rm ecl} = 6400 \Msun$).
Cepheids of higher mass are more likely (by $\approx 30$\%) to be found in clusters than Cepheids of lower mass because lower mass Cepheids appear 
at a later time when more clusters dissolve. 
The evolution of the lower mass ($100 \Msun$) and the more massive cluster ($3200 \Msun$) is shown as videos in the online material.

Clusters orbiting the galaxy at a larger galactocentric radius $R_{\rm g}$ take longer to dissolve in the galactic tidal field, 
retaining more stars (and thus more ProCeps). 
From $R_{\rm g} = 4 \Kpc$ to $R_{\rm g} = 12 \Kpc$, the fraction of Cepheids in clusters $f_{\rm CC}$ increases by a factor of $1.2$. 
At a lower metallicity, Cepheids occur later, at the time when more star clusters are dissolved: as $Z$ changes 
from the Solar value of $0.014$ to a SMC value of $0.002$, $f_{\rm CC}$ decreases by a factor of $1.2$.

At a given time, the majority of star clusters host at most one Cepheid. 
Generally, the probability to observe two and more Cepheids within the same cluster at the same time increases with the cluster mass and age, 
and slightly increases with decreasing metallicity (Figs. \ref{foccur} and \ref{fCCMult}). 
For example, in an initially $6400 \Msun$ cluster at $Z = 0.002$, 
the probability to observe two, three and four Cepheids simultaneously is $7.4$\%, $3.5$\% and $2.5$\%, respectively. 
The highest number of Cepheids at the same time detected in our simulations in any cluster is five, 
indicating that observing more Cepheids in one cluster than this is rare. 
However, we find that the extremely Cepheid-rich LMC clusters NGC\,1866 and NGC\,2103 are consistent with our results once their higher mass is accounted for. Our simulations allow us to understand the existence of such a large number of Cepheids in these clusters as being due to the combination of very high cluster mass, lower LMC metallicity, and a cluster age near the age when the least massive (most numerous) Cepheid progenitors become Cepheids.

Assuming that star clusters are formed with the mass function of \eq{eClusterIMF}, and then evolve in 
a Milky Way-like galaxy, we obtain $f_{\rm CC} = 0.30$ to $0.36$ for the whole galaxy with $Z=0.014$. 
This is an upper limit on $f_{\rm CC}$ because it neglects other cluster dissolving mechanisms than gas expulsion, 
cluster internal dynamics and the galactic tidal field. 
Given that the observed difference of $f_{\rm CC}$ between the galaxy M~31 ($f_{\rm CC} \approx 0.025$) on the one handside, and 
the Galaxy, LMC and SMC ($f_{\rm CC} \approx 0.09$) on the other handside, differs by a substantially larger factor than 
what our simulations permit with a varying tidal field strength and metallicity, it is unlikely that the observed  
difference can be explained by variations of the galactic tidal field strength and metallicity only.

Using an analytical model to incorporate additional cluster disruption mechanisms 
to match the observed number of dissolved clusters \citep{Lada2003}, 
we obtain substantially smaller values of $f_{\rm CC} = 0.015$ to $0.026$. 
This disruption is attributed mainly to interactions with giant molecular clouds, which is not included in 
the present N-body simulations.
Thus, $f_{\rm CC}$ strongly depends on cluster disruption mechanisms, while 
its dependence on the galactic tidal field or metallicity is of secondary importance. 

\begin{acknowledgements}
We thank an anonymous referee for the useful comments and suggestions, which improved the quality of the paper.
We would like to thank Sverre Aarseth for continuously developing the \textsc{nbody} family of numerical integrators 
and Sambaran Banerjee for his advice about specifying integration parameters for the simulations. 
We appreciate the support of the ESO IT team, which was vital for performing the presented simulations.
We thank Mauricio Cruz Reyes for assistance in obtaining an updated value for $f_{\rm CC}$ based on Gaia EDR3 data.
This research made use of Astropy \footnote{http://www.astropy.org}, a community developed core Python package 
for Astronomy \citep{astropy2013} and Matplotlib Python Package \citep{matplotlib2007}.
FD and PK acknowledge the Scientific Visitors Programme of European Southern Observatory in Garching, which made this project possible. 
FD and PK acknowledge support through grant 20-21855S from the Czech Grant Agency.
RIA acknowledges funding provided by SNSF Eccellenza Professorial Fellowship PCEFP2\_194638 
and funding from the European Research Council (ERC) under the European Union's Horizon 2020 research and innovation programme (Grant Agreement No. 947660).
\end{acknowledgements}

%
%




\bibliographystyle{aa} 
\bibliography{cepheidsEscape,richard} 

\begin{thebibliography}{132}
\expandafter\ifx\csname natexlab\endcsname\relax\def\natexlab#1{#1}\fi

\bibitem[{{Aarseth}(1971)}]{Aarseth1971}
{Aarseth}, S.~J. 1971, \apss, 13, 324

\bibitem[{{Aarseth}(1999)}]{Aarseth1999}
{Aarseth}, S.~J. 1999, \pasp, 111, 1333

\bibitem[{{Aarseth}(2003)}]{Aarseth2003}
{Aarseth}, S.~J. 2003, {Gravitational N-Body Simulations} (Cambridge: Cambridge
  University Press)

\bibitem[{{Aarseth} {et~al.}(1974){Aarseth}, {Henon}, \&
  {Wielen}}]{Aarseth1974b}
{Aarseth}, S.~J., {Henon}, M., \& {Wielen}, R. 1974, \aap, 37, 183

\bibitem[{{Aarseth} \& {Zare}(1974)}]{Aarseth1974a}
{Aarseth}, S.~J. \& {Zare}, K. 1974, Celestial Mechanics, 10, 185

\bibitem[{{Ahmad} \& {Cohen}(1973)}]{Ahmad1973}
{Ahmad}, A. \& {Cohen}, L. 1973, Journal of Computational Physics, 12, 389

\bibitem[{{Allen} \& {Santillan}(1991)}]{Allen1991}
{Allen}, C. \& {Santillan}, A. 1991, \rmxaa, 22, 255

\bibitem[{{Anderson} {et~al.}(2013){Anderson}, {Eyer}, \&
  {Mowlavi}}]{Anderson2013}
{Anderson}, R.~I., {Eyer}, L., \& {Mowlavi}, N. 2013, \mnras, 434, 2238

\bibitem[{{Anderson} \& {Riess}(2018)}]{Anderson2018}
{Anderson}, R.~I. \& {Riess}, A.~G. 2018, \apj, 861, 36

\bibitem[{{Anderson} {et~al.}(2016){Anderson}, {Saio}, {Ekstr{\"o}m}, {Georgy},
  \& {Meynet}}]{Anderson2016}
{Anderson}, R.~I., {Saio}, H., {Ekstr{\"o}m}, S., {Georgy}, C., \& {Meynet}, G.
  2016, \aap, 591, A8

\bibitem[{{Astropy Collaboration} {et~al.}(2013){Astropy Collaboration},
  {Robitaille}, {Tollerud}, {Greenfield}, {Droettboom}, {Bray}, {Aldcroft},
  {Davis}, {Ginsburg}, {Price-Whelan}, {Kerzendorf}, {Conley}, {Crighton},
  {Barbary}, {Muna}, {Ferguson}, {Grollier}, {Parikh}, {Nair}, {Unther},
  {Deil}, {Woillez}, {Conseil}, {Kramer}, {Turner}, {Singer}, {Fox}, {Weaver},
  {Zabalza}, {Edwards}, {Azalee Bostroem}, {Burke}, {Casey}, {Crawford},
  {Dencheva}, {Ely}, {Jenness}, {Labrie}, {Lim}, {Pierfederici}, {Pontzen},
  {Ptak}, {Refsdal}, {Servillat}, \& {Streicher}}]{astropy2013}
{Astropy Collaboration}, {Robitaille}, T.~P., {Tollerud}, E.~J., {et~al.} 2013,
  \aap, 558, A33

\bibitem[{{Banerjee} \& {Kroupa}(2013)}]{Banerjee2013}
{Banerjee}, S. \& {Kroupa}, P. 2013, \apj, 764, 29

\bibitem[{{Banerjee} \& {Kroupa}(2017)}]{Banerjee2017}
{Banerjee}, S. \& {Kroupa}, P. 2017, \aap, 597, A28

\bibitem[{{Baumgardt} {et~al.}(2002){Baumgardt}, {Hut}, \&
  {Heggie}}]{Baumgardt2002}
{Baumgardt}, H., {Hut}, P., \& {Heggie}, D.~C. 2002, \mnras, 336, 1069

\bibitem[{{Baumgardt} \& {Kroupa}(2007)}]{Baumgardt2007}
{Baumgardt}, H. \& {Kroupa}, P. 2007, \mnras, 380, 1589

\bibitem[{{Baumgardt} \& {Makino}(2003)}]{Baumgardt2003}
{Baumgardt}, H. \& {Makino}, J. 2003, \mnras, 340, 227

\bibitem[{{Bettis} \& {Szebehely}(1971)}]{Bettis1971}
{Bettis}, D.~G. \& {Szebehely}, V. 1971, \apss, 14, 133

\bibitem[{{Bik} {et~al.}(2003){Bik}, {Lamers}, {Bastian}, {Panagia}, \&
  {Romaniello}}]{Bik2003}
{Bik}, A., {Lamers}, H.~J.~G.~L.~M., {Bastian}, N., {Panagia}, N., \&
  {Romaniello}, M. 2003, \aap, 397, 473

\bibitem[{{Binney} \& {Tremaine}(2008)}]{Binney2008}
{Binney}, J. \& {Tremaine}, S. 2008, {Galactic Dynamics: Second Edition}
  (Princeton University Press)

\bibitem[{{Bonnell} \& {Davies}(1998)}]{Bonnell1998}
{Bonnell}, I.~A. \& {Davies}, M.~B. 1998, \mnras, 295, 691

\bibitem[{{Bono} {et~al.}(2005){Bono}, {Marconi}, {Cassisi}, {Caputo},
  {Gieren}, \& {Pietrzynski}}]{Bono2005}
{Bono}, G., {Marconi}, M., {Cassisi}, S., {et~al.} 2005, \apj, 621, 966

\bibitem[{{Bosma}(1981)}]{Bosma1981}
{Bosma}, A. 1981, \aj, 86, 1825

\bibitem[{{Breuval} {et~al.}(2020){Breuval}, {Kervella}, {Anderson}, {Riess},
  {Arenou}, {Trahin}, {M{\'e}rand}, {Gallenne}, {Gieren}, {Storm}, {Bono},
  {Pietrzy{\'n}ski}, {Nardetto}, {Javanmardi}, \& {Hocd{\'e}}}]{Breuval2020}
{Breuval}, L., {Kervella}, P., {Anderson}, R.~I., {et~al.} 2020, \aap, 643,
  A115

\bibitem[{{Brinkmann} {et~al.}(2017){Brinkmann}, {Banerjee}, {Motwani}, \&
  {Kroupa}}]{Brinkmann2017}
{Brinkmann}, N., {Banerjee}, S., {Motwani}, B., \& {Kroupa}, P. 2017, \aap,
  600, A49

\bibitem[{{Cantat-Gaudin} {et~al.}(2020){Cantat-Gaudin}, {Anders},
  {Castro-Ginard}, {Jordi}, {Romero-G{\'o}mez}, {Soubiran}, {Casamiquela},
  {Tarricq}, {Moitinho}, {Vallenari}, {Bragaglia}, {Krone-Martins}, \&
  {Kounkel}}]{Cantat2020}
{Cantat-Gaudin}, T., {Anders}, F., {Castro-Ginard}, A., {et~al.} 2020, \aap,
  640, A1

\bibitem[{{Cantat-Gaudin} {et~al.}(2018{\natexlab{a}}){Cantat-Gaudin}, {Jordi},
  {Vallenari}, {Bragaglia}, {Balaguer-N{\'u}{\~n}ez}, {Soubiran}, {Bossini},
  {Moitinho}, {Castro-Ginard}, {Krone-Martins}, {Casamiquela}, {Sordo}, \&
  {Carrera}}]{CantatGaudin2018}
{Cantat-Gaudin}, T., {Jordi}, C., {Vallenari}, A., {et~al.} 2018{\natexlab{a}},
  \aap, 618, A93

\bibitem[{{Cantat-Gaudin} {et~al.}(2018{\natexlab{b}}){Cantat-Gaudin}, {Jordi},
  {Vallenari}, {Bragaglia}, {Balaguer-N{\'u}{\~n}ez}, {Soubiran}, {Bossini},
  {Moitinho}, {Castro-Ginard}, {Krone-Martins}, {Casamiquela}, {Sordo}, \&
  {Carrera}}]{Cantat2018}
{Cantat-Gaudin}, T., {Jordi}, C., {Vallenari}, A., {et~al.} 2018{\natexlab{b}},
  \aap, 618, A93

\bibitem[{{Casertano} \& {Hut}(1985)}]{Casertano1985}
{Casertano}, S. \& {Hut}, P. 1985, \apj, 298, 80

\bibitem[{{Castro-Ginard} {et~al.}(2020){Castro-Ginard}, {Jordi}, {Luri},
  {{\'A}lvarez Cid-Fuentes}, {Casamiquela}, {Anders}, {Cantat-Gaudin},
  {Mongui{\'o}}, {Balaguer-N{\'u}{\~n}ez}, {Sol{\`a}}, \&
  {Badia}}]{Castro-Ginard2020}
{Castro-Ginard}, A., {Jordi}, C., {Luri}, X., {et~al.} 2020, \aap, 635, A45

\bibitem[{{Castro-Ginard} {et~al.}(2021){Castro-Ginard}, {Jordi}, {Luri},
  {Cantat-Gaudin}, {Carrasco}, {Casamiquela}, {Anders},
  {Balaguer-N{\'u}{\~n}ez}, \& {Badia}}]{Castro2021}
{Castro-Ginard}, A., {Jordi}, C., {Luri}, X., {et~al.} 2021, arXiv e-prints,
  arXiv:2111.01819

\bibitem[{{de la Fuente Marcos} \& {de la Fuente
  Marcos}(2004)}]{FuenteMarcos2004}
{de la Fuente Marcos}, R. \& {de la Fuente Marcos}, C. 2004, \na, 9, 475

\bibitem[{{D{\'e}k{\'a}ny} {et~al.}(2015){D{\'e}k{\'a}ny}, {Minniti}, {Hajdu},
  {Alonso-Garc{\'\i}a}, {Hempel}, {Palma}, {Catelan}, {Gieren}, \&
  {Majaess}}]{Dekany2015}
{D{\'e}k{\'a}ny}, I., {Minniti}, D., {Hajdu}, G., {et~al.} 2015, \apjl, 799,
  L11

\bibitem[{{Dias} {et~al.}(2021){Dias}, {Monteiro}, {Moitinho}, {L{\'e}pine},
  {Carraro}, {Paunzen}, {Alessi}, \& {Villela}}]{Dias2021}
{Dias}, W.~S., {Monteiro}, H., {Moitinho}, A., {et~al.} 2021, \mnras, 504, 356

\bibitem[{{Dinnbier} \& {Kroupa}(2020)}]{Dinnbier2020a}
{Dinnbier}, F. \& {Kroupa}, P. 2020, \aap, 640, A84

\bibitem[{{Eggleton}(1983)}]{Eggleton1983}
{Eggleton}, P.~P. 1983, \apj, 268, 368

\bibitem[{{Evans} {et~al.}(2020){Evans}, {G{\"u}nther}, {Bond}, {Schaefer},
  {Mason}, {Karovska}, {Tingle}, {Wolk}, {Engle}, {Guinan}, {Pillitteri},
  {Proffitt}, {Kervella}, {Gallenne}, {Anderson}, \& {Moe}}]{Evans2020}
{Evans}, N.~R., {G{\"u}nther}, H.~M., {Bond}, H.~E., {et~al.} 2020, \apj, 905,
  81

\bibitem[{{Ferrarese} {et~al.}(2000){Ferrarese}, {Silbermann}, {Mould},
  {Stetson}, {Saha}, {Freedman}, \& {Kennicutt}}]{Ferrarese2000}
{Ferrarese}, L., {Silbermann}, N.~A., {Mould}, J.~R., {et~al.} 2000, \pasp,
  112, 177

\bibitem[{{Ferreira} {et~al.}(2021){Ferreira}, {Corradi}, {Maia}, {Angelo}, \&
  {Santos}}]{Ferreira2021}
{Ferreira}, F.~A., {Corradi}, W.~J.~B., {Maia}, F.~F.~S., {Angelo}, M.~S., \&
  {Santos}, J.~F.~C., J. 2021, \mnras, 502, L90

\bibitem[{{Fischer} {et~al.}(1992){Fischer}, {Welch}, {Cote}, {Mateo}, \&
  {Madore}}]{Fischer1992}
{Fischer}, P., {Welch}, D.~L., {Cote}, P., {Mateo}, M., \& {Madore}, B.~F.
  1992, \aj, 103, 857

\bibitem[{{Fujii} \& {Portegies Zwart}(2011)}]{Fujii2011}
{Fujii}, M.~S. \& {Portegies Zwart}, S. 2011, Science, 334, 1380

\bibitem[{{Gaia Collaboration} {et~al.}(2021){Gaia Collaboration}, {Brown},
  {Vallenari}, {Prusti}, {de Bruijne}, \& {Babusiaux}}]{Gaiac2021}
{Gaia Collaboration}, {Brown}, A.~G.~A., {Vallenari}, A., {et~al.} 2021, \aap,
  649, A1

\bibitem[{{Gaia Collaboration} {et~al.}(2016){Gaia Collaboration}, {Prusti},
  {de Bruijne}, {Brown}, {Vallenari}, {Babusiaux}, {Bailer-Jones}, {Bastian},
  {Biermann}, {Evans}, \& et~al.}]{Gaiac2016b}
{Gaia Collaboration}, {Prusti}, T., {de Bruijne}, J.~H.~J., {et~al.} 2016,
  \aap, 595, A1

\bibitem[{{Geyer} \& {Burkert}(2001)}]{Geyer2001}
{Geyer}, M.~P. \& {Burkert}, A. 2001, \mnras, 323, 988

\bibitem[{{Gieles} {et~al.}(2006){Gieles}, {Larsen}, {Bastian}, \&
  {Stein}}]{Gieles2006}
{Gieles}, M., {Larsen}, S.~S., {Bastian}, N., \& {Stein}, I.~T. 2006, \aap,
  450, 129

\bibitem[{{Goodwin}(1997)}]{Goodwin1997}
{Goodwin}, S.~P. 1997, \mnras, 284, 785

\bibitem[{{Goodwin} \& {Kroupa}(2005)}]{Goodwin2005}
{Goodwin}, S.~P. \& {Kroupa}, P. 2005, \aap, 439, 565

\bibitem[{{Heggie}(1975)}]{Heggie1975}
{Heggie}, D.~C. 1975, \mnras, 173, 729

\bibitem[{{Holl} {et~al.}(2018){Holl}, {Audard}, {Nienartowicz}, {Jevardat de
  Fombelle}, {Marchal}, {Mowlavi}, {Clementini}, {De Ridder}, {Evans}, {Guy},
  {Lanzafame}, {Lebzelter}, {Rimoldini}, {Roelens}, {Zucker}, {Distefano},
  {Garofalo}, {Lecoeur-Ta{\"\i}bi}, {Lopez}, {Molinaro}, {Muraveva}, {Panahi},
  {Regibo}, {Ripepi}, {Sarro}, {Aerts}, {Anderson}, {Charnas}, {Barblan},
  {Blanco-Cuaresma}, {Busso}, {Cuypers}, {De Angeli}, {Glass}, {Grenon},
  {Juh{\'a}sz}, {Kochoska}, {Koubsky}, {Lanza}, {Leccia}, {Lorenz}, {Marconi},
  {Marschalk{\'o}}, {Mazeh}, {Messina}, {Mignard}, {Moitinho}, {Moln{\'a}r},
  {Morgenthaler}, {Musella}, {Ordenovic}, {Ord{\'o}{\~n}ez}, {Pagano},
  {Palaversa}, {Pawlak}, {Plachy}, {Pr{\v{s}}a}, {Riello}, {S{\"u}veges},
  {Szabados}, {Szegedi-Elek}, {Votruba}, \& {Eyer}}]{GaiaDR2variability}
{Holl}, B., {Audard}, M., {Nienartowicz}, K., {et~al.} 2018, \aap, 618, A30

\bibitem[{{Hunt} \& {Reffert}(2021)}]{Hunt2021}
{Hunt}, E.~L. \& {Reffert}, S. 2021, \aap, 646, A104

\bibitem[{Hunter(2007)}]{matplotlib2007}
Hunter, J.~D. 2007, Computing in Science \& Engineering, 9, 90

\bibitem[{{Hurley} {et~al.}(2000){Hurley}, {Pols}, \& {Tout}}]{Hurley2000}
{Hurley}, J.~R., {Pols}, O.~R., \& {Tout}, C.~A. 2000, \mnras, 315, 543

\bibitem[{{Hurley} {et~al.}(2004){Hurley}, {Tout}, {Aarseth}, \&
  {Pols}}]{Hurley2004}
{Hurley}, J.~R., {Tout}, C.~A., {Aarseth}, S.~J., \& {Pols}, O.~R. 2004,
  \mnras, 355, 1207

\bibitem[{{Hurley} {et~al.}(2002){Hurley}, {Tout}, \& {Pols}}]{Hurley2002}
{Hurley}, J.~R., {Tout}, C.~A., \& {Pols}, O.~R. 2002, \mnras, 329, 897

\bibitem[{{Irrgang} {et~al.}(2013){Irrgang}, {Wilcox}, {Tucker}, \&
  {Schiefelbein}}]{Irrgang2013}
{Irrgang}, A., {Wilcox}, B., {Tucker}, E., \& {Schiefelbein}, L. 2013, \aap,
  549, A137

\bibitem[{{Irwin}(1955)}]{Irwin1955}
{Irwin}, J.~B. 1955, Monthly Notes of the Astronomical Society of South Africa,
  14, 38

\bibitem[{{Jerabkova} {et~al.}(2021){Jerabkova}, {Boffin}, {Beccari}, {de
  Marchi}, {de Bruijne}, \& {Prusti}}]{Jerabkova2021}
{Jerabkova}, T., {Boffin}, H. M.~J., {Beccari}, G., {et~al.} 2021, \aap, 647,
  A137

\bibitem[{{Johnson} {et~al.}(2017){Johnson}, {Seth}, {Dalcanton}, {Beerman},
  {Fouesneau}, {Weisz}, {Bell}, {Dolphin}, {Sandstrom}, \&
  {Williams}}]{Johnson2017}
{Johnson}, L.~C., {Seth}, A.~C., {Dalcanton}, J.~J., {et~al.} 2017, \apj, 839,
  78

\bibitem[{{Kholopov}(1956)}]{Kholopov1956}
{Kholopov}, P.~N. 1956, Peremennye Zvezdy, 11, 325

\bibitem[{{Kochanek}(1992)}]{Kochanek1992}
{Kochanek}, C.~S. 1992, \apj, 385, 604

\bibitem[{{Kroupa}(1995{\natexlab{a}})}]{Kroupa1995a}
{Kroupa}, P. 1995{\natexlab{a}}, \mnras, 277, 1491

\bibitem[{{Kroupa}(1995{\natexlab{b}})}]{Kroupa1995b}
{Kroupa}, P. 1995{\natexlab{b}}, \mnras, 277, 1507

\bibitem[{{Kroupa}(2001)}]{Kroupa2001a}
{Kroupa}, P. 2001, \mnras, 322, 231

\bibitem[{{Kroupa} {et~al.}(2001){Kroupa}, {Aarseth}, \&
  {Hurley}}]{Kroupa2001b}
{Kroupa}, P., {Aarseth}, S., \& {Hurley}, J. 2001, \mnras, 321, 699

\bibitem[{{Kuhn} {et~al.}(2014){Kuhn}, {Feigelson}, {Getman}, {Baddeley},
  {Broos}, {Sills}, {Bate}, {Povich}, {Luhman}, {Busk}, {Naylor}, \&
  {King}}]{Kuhn2014}
{Kuhn}, M.~A., {Feigelson}, E.~D., {Getman}, K.~V., {et~al.} 2014, \apj, 787,
  107

\bibitem[{{K{\"u}pper} {et~al.}(2010){K{\"u}pper}, {Kroupa}, {Baumgardt}, \&
  {Heggie}}]{Kupper2010}
{K{\"u}pper}, A.~H.~W., {Kroupa}, P., {Baumgardt}, H., \& {Heggie}, D.~C. 2010,
  \mnras, 401, 105

\bibitem[{{K{\"u}pper} {et~al.}(2008){K{\"u}pper}, {MacLeod}, \&
  {Heggie}}]{Kupper2008}
{K{\"u}pper}, A.~H.~W., {MacLeod}, A., \& {Heggie}, D.~C. 2008, \mnras, 387,
  1248

\bibitem[{{K{\"u}pper} {et~al.}(2011){K{\"u}pper}, {Maschberger}, {Kroupa}, \&
  {Baumgardt}}]{Kupper2011}
{K{\"u}pper}, A. H.~W., {Maschberger}, T., {Kroupa}, P., \& {Baumgardt}, H.
  2011, \mnras, 417, 2300

\bibitem[{{Kustaanheimo} \& {Stiefel}(1965)}]{Kustaanheimo1965}
{Kustaanheimo}, P. \& {Stiefel}, E. 1965, Reine Angew. Math., 218, 204

\bibitem[{{Lada} \& {Lada}(2003)}]{Lada2003}
{Lada}, C.~J. \& {Lada}, E.~A. 2003, \araa, 41, 57

\bibitem[{{Lada} {et~al.}(1984){Lada}, {Margulis}, \& {Dearborn}}]{Lada1984}
{Lada}, C.~J., {Margulis}, M., \& {Dearborn}, D. 1984, \apj, 285, 141

\bibitem[{{Lada} {et~al.}(1991){Lada}, {Depoy}, {Evans}, \&
  {Gatley}}]{Lada1991}
{Lada}, E.~A., {Depoy}, D.~L., {Evans}, Neal~J., I., \& {Gatley}, I. 1991,
  \apj, 371, 171

\bibitem[{{Leavitt} \& {Pickering}(1912)}]{Leavitt1912}
{Leavitt}, H.~S. \& {Pickering}, E.~C. 1912, Harvard College Observatory
  Circular, 173, 1

\bibitem[{{Lewis} {et~al.}(2015){Lewis}, {Dolphin}, {Dalcanton}, {Weisz},
  {Williams}, {Bell}, {Seth}, {Simones}, {Skillman}, {Choi}, {Fouesneau},
  {Guhathakurta}, {Johnson}, {Kalirai}, {Leroy}, {Monachesi}, {Rix}, \&
  {Schruba}}]{Lewis2015}
{Lewis}, A.~R., {Dolphin}, A.~E., {Dalcanton}, J.~J., {et~al.} 2015, \apj, 805,
  183

\bibitem[{{Loktin} \& {Popova}(2017)}]{Loktin2017}
{Loktin}, A.~V. \& {Popova}, M.~E. 2017, Astrophysical Bulletin, 72, 257

\bibitem[{{Majaess} {et~al.}(2013){Majaess}, {Carraro}, {Moni Bidin},
  {Bonatto}, {Berdnikov}, {Balam}, {Moyano}, {Gallo}, {Turner}, {Lane},
  {Gieren}, {Borissova}, {Kovtyukh}, \& {Beletsky}}]{Majaess2013}
{Majaess}, D., {Carraro}, G., {Moni Bidin}, C., {et~al.} 2013, \aap, 560, A22

\bibitem[{{Makino}(1991)}]{Makino1991}
{Makino}, J. 1991, \apj, 369, 200

\bibitem[{{Makino} \& {Aarseth}(1992)}]{Makino1992}
{Makino}, J. \& {Aarseth}, S.~J. 1992, \pasj, 44, 141

\bibitem[{{Mardling} \& {Aarseth}(1999)}]{Mardling1999}
{Mardling}, R. \& {Aarseth}, S. 1999, in NATO Advanced Study Institute (ASI)
  Series C, Vol. 522, The Dynamics of Small Bodies in the Solar System, A Major
  Key to Solar System Studies, ed. B.~A. {Steves} \& A.~E. {Roy}, 385

\bibitem[{{Marks} \& {Kroupa}(2012)}]{Marks2012}
{Marks}, M. \& {Kroupa}, P. 2012, \aap, 543, A8

\bibitem[{{McMillan} {et~al.}(2007){McMillan}, {Vesperini}, \& {Portegies
  Zwart}}]{McMillan2007}
{McMillan}, S. L.~W., {Vesperini}, E., \& {Portegies Zwart}, S.~F. 2007, \apjl,
  655, L45

\bibitem[{{Medina} {et~al.}(2021){Medina}, {Lemasle}, \& {Grebel}}]{Medina2021}
{Medina}, G.~E., {Lemasle}, B., \& {Grebel}, E.~K. 2021, \mnras, 505, 1342

\bibitem[{{Megeath} {et~al.}(2016){Megeath}, {Gutermuth}, {Muzerolle},
  {Kryukova}, {Hora}, {Allen}, {Flaherty}, {Hartmann}, {Myers}, {Pipher},
  {Stauffer}, {Young}, \& {Fazio}}]{Megeath2016}
{Megeath}, S.~T., {Gutermuth}, R., {Muzerolle}, J., {et~al.} 2016, \aj, 151, 5

\bibitem[{{Meingast} \& {Alves}(2019)}]{Meingast2019}
{Meingast}, S. \& {Alves}, J. 2019, \aap, 621, L3

\bibitem[{{Mikkola} \& {Aarseth}(1990)}]{Mikkola1990}
{Mikkola}, S. \& {Aarseth}, S.~J. 1990, Celestial Mechanics and Dynamical
  Astronomy, 47, 375

\bibitem[{{Miyamoto} \& {Nagai}(1975)}]{Miyamoto1975}
{Miyamoto}, M. \& {Nagai}, R. 1975, \pasj, 27, 533

\bibitem[{{Moe} \& {Di Stefano}(2017)}]{Moe2017}
{Moe}, M. \& {Di Stefano}, R. 2017, \apjs, 230, 15

\bibitem[{{Mouri} \& {Taniguchi}(2002)}]{Mouri2002}
{Mouri}, H. \& {Taniguchi}, Y. 2002, \apj, 580, 844

\bibitem[{{Musella} {et~al.}(2016){Musella}, {Marconi}, {Stetson}, {Raimondo},
  {Brocato}, {Molinaro}, {Ripepi}, {Carini}, {Coppola}, {Walker}, \&
  {Welch}}]{Musella2016}
{Musella}, I., {Marconi}, M., {Stetson}, P.~B., {et~al.} 2016, \mnras, 457,
  3084

\bibitem[{{Oh} \& {Kroupa}(2016)}]{Oh2016}
{Oh}, S. \& {Kroupa}, P. 2016, \aap, 590, A107

\bibitem[{{Oh} {et~al.}(2015){Oh}, {Kroupa}, \& {Pflamm-Altenburg}}]{Oh2015}
{Oh}, S., {Kroupa}, P., \& {Pflamm-Altenburg}, J. 2015, \apj, 805, 92

\bibitem[{{Pawlak} {et~al.}(2019){Pawlak}, {Pejcha}, {Jakub{\v{c}}{\'\i}k},
  {Jayasinghe}, {Kochanek}, {Stanek}, {Shappee}, {Holoien}, {Thompson},
  {Prieto}, {Dong}, {Shields}, {Pojmanski}, {Britt}, \&
  {Will}}]{ASASSNvarstar2019}
{Pawlak}, M., {Pejcha}, O., {Jakub{\v{c}}{\'\i}k}, P., {et~al.} 2019, \mnras,
  487, 5932

\bibitem[{{Perets} \& {{\v S}ubr}(2012)}]{Perets2012}
{Perets}, H.~B. \& {{\v S}ubr}, L. 2012, \apj, 751, 133

\bibitem[{{Pflamm-Altenburg} {et~al.}(2013){Pflamm-Altenburg},
  {Gonz{\'a}lez-L{\'o}pezlira}, \& {Kroupa}}]{PflammAltenburg2013}
{Pflamm-Altenburg}, J., {Gonz{\'a}lez-L{\'o}pezlira}, R.~A., \& {Kroupa}, P.
  2013, \mnras, 435, 2604

\bibitem[{{Pflamm-Altenburg} \& {Kroupa}(2006)}]{PflammAltenburg2006}
{Pflamm-Altenburg}, J. \& {Kroupa}, P. 2006, \mnras, 373, 295

\bibitem[{{Pflamm-Altenburg} \& {Kroupa}(2008)}]{PflammAltenburg2008}
{Pflamm-Altenburg}, J. \& {Kroupa}, P. 2008, \nat, 455, 641

\bibitem[{{Porras} {et~al.}(2003){Porras}, {Christopher}, {Allen}, {Di
  Francesco}, {Megeath}, \& {Myers}}]{Porras2003}
{Porras}, A., {Christopher}, M., {Allen}, L., {et~al.} 2003, \aj, 126, 1916

\bibitem[{{Pouliasis} {et~al.}(2017){Pouliasis}, {Di Matteo}, \&
  {Haywood}}]{Pouliasis2017}
{Pouliasis}, E., {Di Matteo}, P., \& {Haywood}, M. 2017, \aap, 598, A66

\bibitem[{{Riess} {et~al.}(2020){Riess}, {Yuan}, {Casertano}, {Macri}, \&
  {Scolnic}}]{Riess2020}
{Riess}, A.~G., {Yuan}, W., {Casertano}, S., {Macri}, L.~M., \& {Scolnic}, D.
  2020, \apjl, 896, L43

\bibitem[{{Ripepi} {et~al.}(2019){Ripepi}, {Molinaro}, {Musella}, {Marconi},
  {Leccia}, \& {Eyer}}]{Ripepi2019}
{Ripepi}, V., {Molinaro}, R., {Musella}, I., {et~al.} 2019, \aap, 625, A14

\bibitem[{{R{\"o}ser} {et~al.}(2019){R{\"o}ser}, {Schilbach}, \&
  {Goldman}}]{Roser2019a}
{R{\"o}ser}, S., {Schilbach}, E., \& {Goldman}, B. 2019, \aap, 621, L2

\bibitem[{{Rubin} {et~al.}(1982){Rubin}, {Ford}, {Thonnard}, \&
  {Burstein}}]{Rubin1982}
{Rubin}, V.~C., {Ford}, W.~K., J., {Thonnard}, N., \& {Burstein}, D. 1982,
  \apj, 261, 439

\bibitem[{{Samus'} {et~al.}(2017){Samus'}, {Kazarovets}, {Durlevich},
  {Kireeva}, \& {Pastukhova}}]{GCVS2017}
{Samus'}, N.~N., {Kazarovets}, E.~V., {Durlevich}, O.~V., {Kireeva}, N.~N., \&
  {Pastukhova}, E.~N. 2017, Astronomy Reports, 61, 80

\bibitem[{{Sana} {et~al.}(2012){Sana}, {de Mink}, {de Koter}, {Langer},
  {Evans}, {Gieles}, {Gosset}, {Izzard}, {Le Bouquin}, \&
  {Schneider}}]{Sana2012}
{Sana}, H., {de Mink}, S.~E., {de Koter}, A., {et~al.} 2012, Science, 337, 444

\bibitem[{{Sandage}(1958)}]{Sandage1958}
{Sandage}, A. 1958, \apj, 128, 150

\bibitem[{{Scargle} {et~al.}(2013){Scargle}, {Norris}, {Jackson}, \&
  {Chiang}}]{Scargle2013}
{Scargle}, J.~D., {Norris}, J.~P., {Jackson}, B., \& {Chiang}, J. 2013, \apj,
  764, 167

\bibitem[{{Schoettler} {et~al.}(2020){Schoettler}, {de Bruijne}, {Vaher}, \&
  {Parker}}]{Schoettler2020}
{Schoettler}, C., {de Bruijne}, J., {Vaher}, E., \& {Parker}, R.~J. 2020,
  \mnras, 495, 3104

\bibitem[{{Senchyna} {et~al.}(2015){Senchyna}, {Johnson}, {Dalcanton},
  {Beerman}, {Fouesneau}, {Dolphin}, {Williams}, {Rosenfield}, \&
  {Larsen}}]{Senchyna2015}
{Senchyna}, P., {Johnson}, L.~C., {Dalcanton}, J.~J., {et~al.} 2015, \apj, 813,
  31

\bibitem[{{Sollima}(2020)}]{Sollima2020}
{Sollima}, A. 2020, \mnras, 495, 2222

\bibitem[{{Soszy{\'n}ski} {et~al.}(2020){Soszy{\'n}ski}, {Udalski},
  {Szyma{\'n}ski}, {Pietrukowicz}, {Skowron}, {Skowron}, {Poleski},
  {Koz{\l}owski}, {Mr{\'o}z}, {Ulaczyk}, {Rybicki}, {Iwanek}, {Wrona}, \&
  {Gromadzki}}]{Soszynski2020galcep}
{Soszy{\'n}ski}, I., {Udalski}, A., {Szyma{\'n}ski}, M.~K., {et~al.} 2020,
  \actaa, 70, 101

\bibitem[{{Spera} {et~al.}(2016){Spera}, {Mapelli}, \& {Jeffries}}]{Spera2016}
{Spera}, M., {Mapelli}, M., \& {Jeffries}, R.~D. 2016, \mnras, 460, 317

\bibitem[{{Spitzer}(1958)}]{Spitzer1958}
{Spitzer}, Lyman, J. 1958, \apj, 127, 17

\bibitem[{{Spitzer}(1969)}]{Spitzer1969}
{Spitzer}, Jr., L. 1969, \apjl, 158, L139

\bibitem[{{Stanimirovi{\'c}} {et~al.}(2004){Stanimirovi{\'c}},
  {Staveley-Smith}, \& {Jones}}]{Stanimirovic2004}
{Stanimirovi{\'c}}, S., {Staveley-Smith}, L., \& {Jones}, P.~A. 2004, \apj,
  604, 176

\bibitem[{{Sterzik} \& {Durisen}(1998)}]{Sterzik1998}
{Sterzik}, M.~F. \& {Durisen}, R.~H. 1998, \aap, 339, 95

\bibitem[{{Stiefel} \& {Scheifele}(1975)}]{Stiefel1975}
{Stiefel}, E.~L. \& {Scheifele}, G. 1975, {Linear and regular celestial
  mechanics. Perturbed two-body motion. Numerical methods. Canonical theory.}

\bibitem[{{Tanikawa} {et~al.}(2012){Tanikawa}, {Hut}, \&
  {Makino}}]{Tanikawa2012}
{Tanikawa}, A., {Hut}, P., \& {Makino}, J. 2012, \na, 17, 272

\bibitem[{{Terlevich}(1987)}]{Terlevich1987}
{Terlevich}, E. 1987, \mnras, 224, 193

\bibitem[{{Testa} {et~al.}(2007){Testa}, {Marconi}, {Musella}, {Ripepi},
  {Dall'Ora}, {Ferraro}, {Mucciarelli}, {Mateo}, \& {C{\^o}t{\'e}}}]{Testa2007}
{Testa}, V., {Marconi}, M., {Musella}, I., {et~al.} 2007, \aap, 462, 599

\bibitem[{{Theuns}(1992)}]{Theuns1992}
{Theuns}, T. 1992, \aap, 259, 493

\bibitem[{{Tout} {et~al.}(1996){Tout}, {Pols}, {Eggleton}, \& {Han}}]{Tout1996}
{Tout}, C.~A., {Pols}, O.~R., {Eggleton}, P.~P., \& {Han}, Z. 1996, \mnras,
  281, 257

\bibitem[{{Turner}(1976)}]{Turner1976}
{Turner}, D.~G. 1976, \aj, 81, 1125

\bibitem[{{Turner} \& {Burke}(2002)}]{Turner2002}
{Turner}, D.~G. \& {Burke}, J.~F. 2002, \aj, 124, 2931

\bibitem[{{Tutukov}(1978)}]{Tutukov1978}
{Tutukov}, A.~V. 1978, \aap, 70, 57

\bibitem[{{van der Marel} \& {Kallivayalil}(2014)}]{Marel2014}
{van der Marel}, R.~P. \& {Kallivayalil}, N. 2014, \apj, 781, 121

\bibitem[{{Vesperini} \& {Heggie}(1997)}]{Vesperini1997}
{Vesperini}, E. \& {Heggie}, D.~C. 1997, \mnras, 289, 898

\bibitem[{{Wang} {et~al.}(2019){Wang}, {Kroupa}, \& {Jerabkova}}]{Wang2019}
{Wang}, L., {Kroupa}, P., \& {Jerabkova}, T. 2019, \mnras, 484, 1843

\bibitem[{{Watson} {et~al.}(2006){Watson}, {Henden}, \& {Price}}]{VSX}
{Watson}, C.~L., {Henden}, A.~A., \& {Price}, A. 2006, Society for Astronomical
  Sciences Annual Symposium, 25, 47

\bibitem[{{Weidner} {et~al.}(2010){Weidner}, {Kroupa}, \&
  {Bonnell}}]{Weidner2010}
{Weidner}, C., {Kroupa}, P., \& {Bonnell}, I.~A.~D. 2010, \mnras, 401, 275

\bibitem[{{Weidner} {et~al.}(2004){Weidner}, {Kroupa}, \&
  {Larsen}}]{Weidner2004}
{Weidner}, C., {Kroupa}, P., \& {Larsen}, S.~S. 2004, \mnras, 350, 1503

\bibitem[{{Welch} \& {Stetson}(1993)}]{Welch1993}
{Welch}, D.~L. \& {Stetson}, P.~B. 1993, \aj, 105, 1813

\bibitem[{{Whitmore} {et~al.}(1999){Whitmore}, {Zhang}, {Leitherer}, {Fall},
  {Schweizer}, \& {Miller}}]{Whitmore1999}
{Whitmore}, B.~C., {Zhang}, Q., {Leitherer}, C., {et~al.} 1999, \aj, 118, 1551

\bibitem[{{Zhou} \& {Chen}(2021)}]{Zhou2021}
{Zhou}, X. \& {Chen}, X. 2021, \mnras, 504, 4768

\end{thebibliography}

\begin{appendix}
\section{Detection of Cepheids originating from stellar mergers}

\label{saMergers}

Since we detect Cepheids automatically according to their evolution in the HR diagram 
(the description is in \refs{sIdentification}), 
it is important to make sure that the recipe is able to deal not only with stars evolving in 
isolation, but also with more complicated cases including binary star evolution and stellar mergers. 
Typical examples (for $Z=0.014$) are shown in \reff{fHRAdditional}. 

The left panel details the evolution of a $11 \Msun$ and $7 \Msun$ binary with initial orbital period of $27$ days, 
which coalesces when the primary increases in radius as it evolves out of the MS (the discontinuity before 
core He burning). 
The merger mass is $14.1 \Msun$. 
Core He burning (orange points) then commences leftwards from the blue boundary of the instability strip, and the star 
monotonically decreases in effective temperature during this stage, which is dissimilar to the loop experienced by Cepheids. 
For this reason, the star does not fulfil condition (iii) for detection of Cepheids (\refs{sIdentification}). 

The middle panel shows a $8.4 \Msun$ primary with a secondary of $1.1 \Msun$ and initial orbital period of $13$ days. 
In this case, the coalescence increases the mass only slightly, and the ProCep becomes a Cepheid. 

The right panel shows coalescence of a primary of ZAMS mass $4.3 \Msun$ with its $1.6 \Msun$ companion. 
For metallicity $Z=0.014$, the initial mass of the primary is lower than 
the mass threshold for Cepheids, which is $m_{\rm min,Ceph} = 4.7 \Msun$. 
After the coalescence, the merged star has a mass of $5.3 \Msun$, and it becomes a Cepheid. 

The cases in the middle and right panel are detected as Cepheids by the automatic routine. 
The case in the left panel is not detected as a Cepheid. 
This demonstrates that the automatic procedure is able to identify Cepheids also 
for more complicated evolutionary tracks including mergers.

\iffigscl
\begin{figure*}
\includegraphics[width=\textwidth]{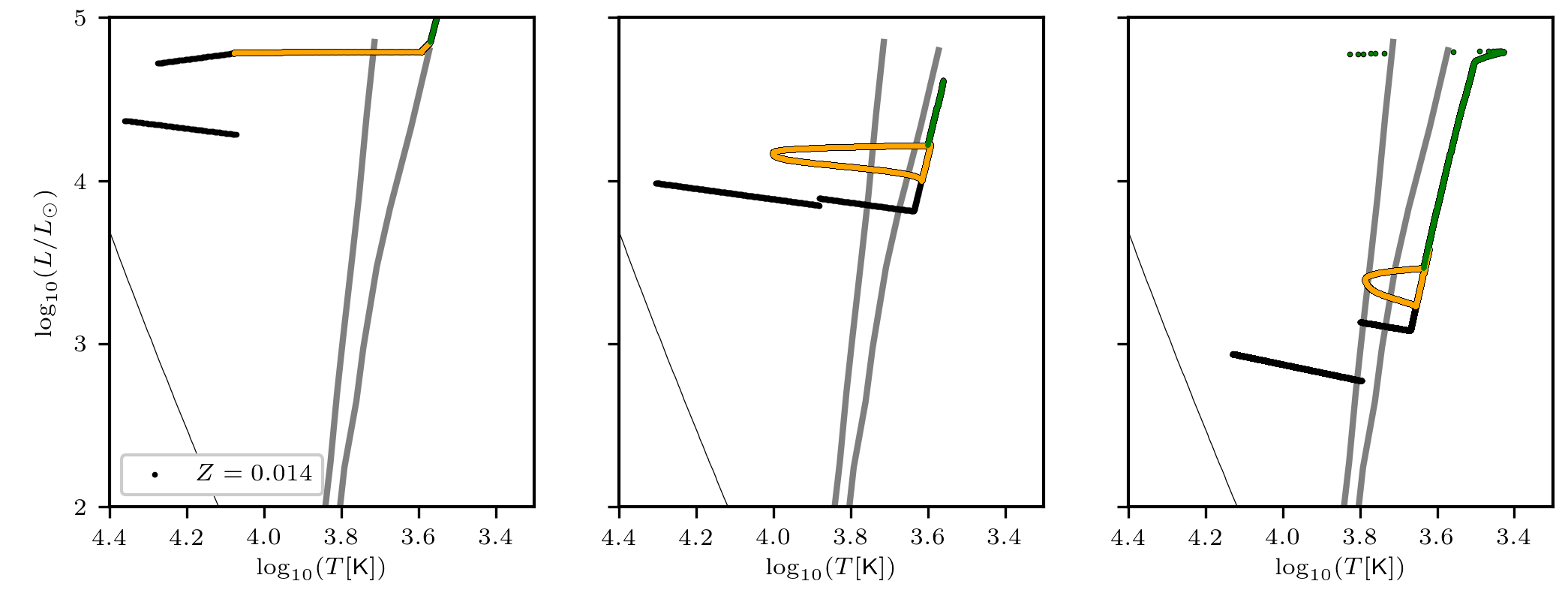}
\caption{
Examples of HR diagrams where the evolutionary path of the primary had been altered by 
a merger with the secondary. 
These plots show examples from the library of simulations calculated by \nbdvi for $Z=0.014$. 
The evolution from the terminal age main-sequence to core He burning is represented by black points, during core He burning by orange points, 
and afterwards by green points. 
The coalescence causes a sudden increase of luminosity, which can be seen as the discontinuity in the evolutionary tracks. 
The instability strip and ZAMS are shown by the thick and thin black lines, respectively. 
\figpan{Left panel}: Coalescence with relatively massive secondary ($7 \Msun$), which 
increases the primary mass so much that the merger object does not become a Cepheid. 
\figpan{Middle panel}: Coalescence with a lower mass secondary, which does not prevent the star 
from becoming a Cepheid. 
\figpan{Right panel}: 
Lower mass primary ($4.3 \Msun$), which would not have evolved to a Cepheid in isolation, 
becomes a Cepheid after a coalescence. 
}
\label{fHRAdditional}
\end{figure*} \else \fi

\section{Comparison of the evolution of ProCeps in a lower mass ($100 \Msun$) and a more massive ($3200 \Msun$) cluster}

\label{saLowHighMassCluster}

The lower mass cluster, which contains three ProCeps in total, segregates its more massive stars before the onset of gas 
expulsion at $0.6 \Myr$ (upper row of \reff{fLowHighMassCluster}). 
At this time (upper middle panel), two ProCeps are already near the cluster centre. 
The third ProCep, which is located at $x \approx -0.6 \Pc$, returns to the cluster centre, where it interacts with the other two Cepheids forming a 
densely bound system. 
This results in all three ProCeps being concentrated at the cluster centre, while lower mass stars are more dispersed or escaping the cluster (upper right panel). 
The complicated interaction between the most massive bodies ejects one of the ProCep and a lower mass star at $7.2 \Myr$ (shown only in the online material). 

In contrast, the more massive star cluster (lower row of \reff{fLowHighMassCluster}) does not mass segregate ProCeps before gas expulsion, which results in the 
population of ProCeps to be dispersed throughout the cluster to a similar degree as the lower mass stars. 
Gas expulsion then unbinds a comparable fraction of ProCeps as low mass stars from the cluster.

\iffigscl
\begin{figure*}
\includegraphics[width=\textwidth]{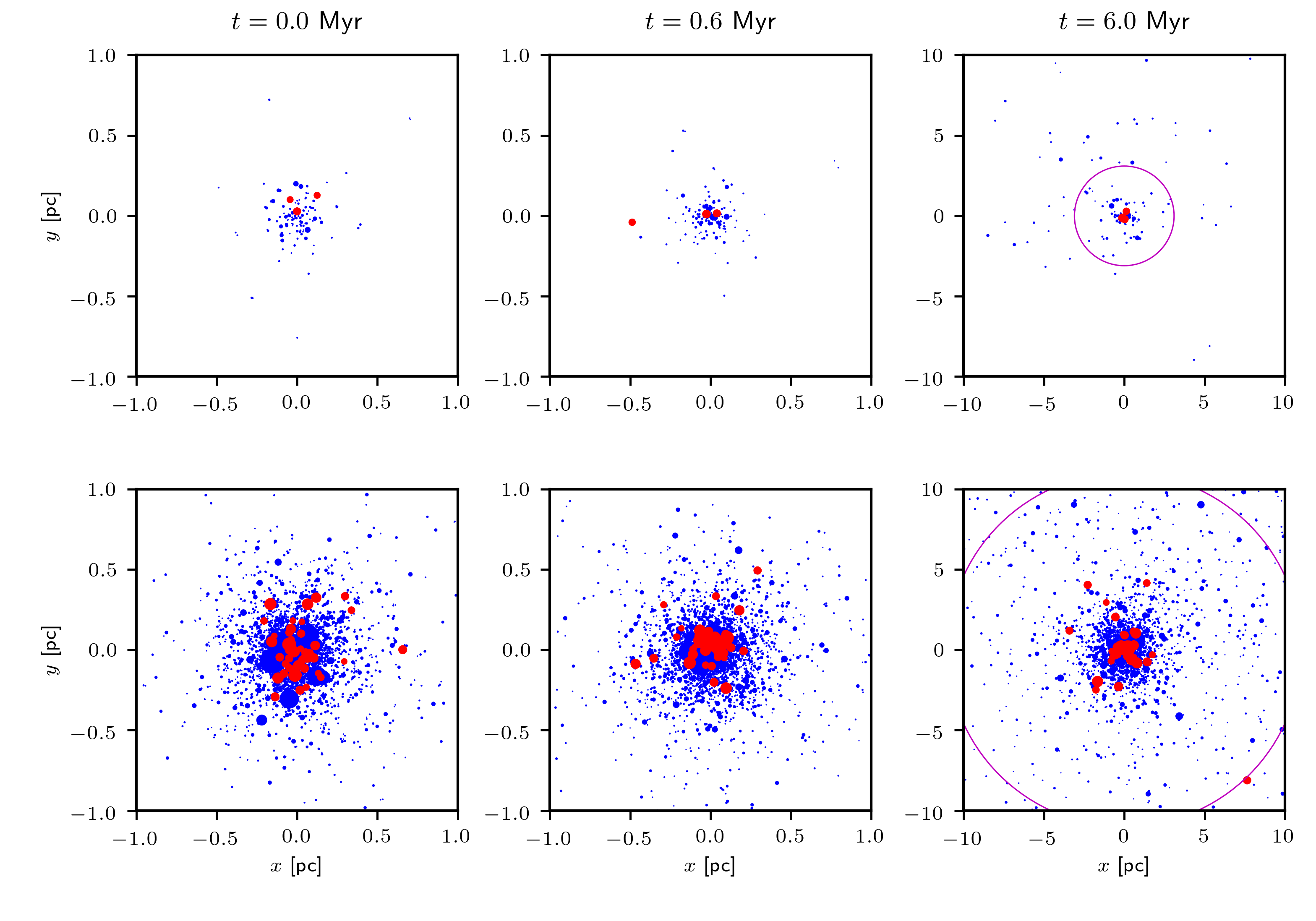}
\caption{
Evolutionary snapshots for an example of a lower mass ($100 \Msun$; upper row) and more massive star clusters ($3200 \Msun$; lower row). 
The plots are shown at the beginning of the simulations (left column), immediately before the beginning of gas expulsion (middle column), 
and after gas expulsion; the time is indicated above the panels. 
ProCeps are marked red, while stars with mass below $m_{\rm min,Ceph}$ or above $m_{\rm max,Ceph}$ are marked blue. 
Compact objects (seen only in the online material) are marked black.
Stellar mass is proportional to the area of the dots.
The magenta circle shows the cluster tidal radius. 
Note that the length-scale on the right column is inflated by a factor of $10$.
The evolution of both star clusters up to $12 \Myr$ can be seen as movies in the online material.
}
\label{fLowHighMassCluster}
\end{figure*} \else \fi

\end{appendix}

\end{document}